\documentclass[journal]{IEEEtran}
\usepackage{psfrag}
\usepackage{amsmath}
\usepackage{siunitx}
\usepackage{amssymb}
\usepackage{color}
\usepackage{graphicx}
\usepackage{epsfig}
\usepackage{fancybox}
\usepackage[FIGTOPCAP]{subfigure}
\usepackage{cite}
\usepackage{algorithm}
\usepackage{algorithmicx}
\usepackage{algpseudocode}
\usepackage{multirow}
\usepackage{theorem}
\usepackage{epstopdf}	
\usepackage{color}		
\usepackage{multicol}
\usepackage{url}
\usepackage{multirow}
\usepackage[table]{xcolor}
\definecolor{light-gray}{gray}{0.95}
\definecolor{darker-gray}{gray}{0.5}

\newcommand{\ignore}[1]{}

\usepackage{xcolor,colortbl}

\definecolor{Gray}{gray}{0.85}
\definecolor{LightCyan}{rgb}{0.88,1,1}

\newcolumntype{a}{>{\columncolor{Gray}}|c|}
\newcolumntype{b}{>{\columncolor{white}}|c|}

\usepackage[latin1]{inputenc}
\usepackage{tikz}
\usetikzlibrary{shapes,arrows}

\tikzstyle{decision} = [diamond, draw, fill=blue!20, 
    text width=4.5em, text badly centered, node distance=3cm, inner sep=0pt]
\tikzstyle{block} = [rectangle, draw, fill=blue!20, 
    text width=5em, text centered, rounded corners, minimum height=4em]
\tikzstyle{line} = [draw, -latex']
\tikzstyle{cloud} = [draw, ellipse,fill=red!20, node distance=3cm,
    minimum height=2em]

\begin{document}

\title{STEM: A Scheme for Two-phase \\
Evaluation of Majority Logic}

\author{Meghna G. Mankalale, Zhaoxin Liang, and Sachin S. Sapatnekar\thanks{The authors are with
the Department of Electrical and Computer Engineering, University of Minnesota,
Minneapolis, MN 55455, USA.  This work was supported in part by C-SPIN, one of
the six SRC STARnet Centers, sponsored by MARCO and DARPA. Copyright
(c) 2017 IEEE. Personal use of this material is permitted. However,
permission to use this material for any other purposes must be
obtained from the IEEE by sending a request to
pubs-permissions@ieee.org.}}

%

\maketitle
\begin{abstract}
The switching time of a magnet in a spin current based majority gate depends on
the input vector combination, and this often restricts the speed of
majority-based circuits. To address this issue, this work proposes a novel
two-phase scheme to implement majority logic and evaluates it on an all-spin
logic (ASL) majority-based logic structures. In Phase 1, the output is
initialized to a preset value. Next in Phase 2, the inputs are evaluated to
switch the output magnet to its correct value. The time window for the output
to switch in Phase 2 is fixed. Using such a scheme, an $n$-input AND gate
which requires a total of ($2n-1$) inputs in the conventional implementation can
now be implemented with only ($n+1$) inputs.  When applied to standard logic
functions, it is demonstrated that the proposed method of designing ASL gates 
are 1.6--3.4$\times$ faster and 1.9--6.9$\times$ more energy-efficient than the 
conventional method, and for a five-magnet full adder, it is shown that
the proposed ASL implementation is 1.5$\times$ faster, 2.2$\times$
more energy-efficient, and provides a 16\% improvement in area.
\end{abstract}


\IEEEpeerreviewmaketitle

\section{Introduction}
\label{sec:intro}

Spintronics is widely viewed as a promising technology for the post-CMOS
era~\cite{BeyondCMOS}. Several new concepts for implementing logic functions
using spin-based devices have been proposed in this regard, including quantum
cellular automata (QCA)~\cite{QCA1, QCA2, QCA3}, all-spin logic
(ASL)~\cite{behin2010proposal}, charge spin logic~\cite{CSL}, and
magnetoelectric-based logic~\cite{IntelME, Mankalale16}. The fundamental logic
building blocks implemented with these devices are based on the majority logic
paradigm, where the input states compete with each other and the majority
prevails as the output logic state.  In this work, we propose a scheme that
speeds up the implementation of majority-logic-based devices using a two-phase
method, and we demonstrate this idea with the help of ASL gates.  

The simplest ASL gate is an inverter~\cite{Kim2015}, shown in Fig.~\ref{fig:ASL}, which
consists of a ferromagnet at the input and another at the output, with a
non-magnetic channel between the two. An applied voltage $V_{dd}$ on the input
magnet results in electron flow due to charge current from $Gnd$ to $V_{dd}$.
The input magnet polarizes the charge current and generates a spin current. The
spins that align with the direction of the magnetization pass through the
magnet and those that are opposite to it are reflected onto the channel. This
creates an accumulation of electron spins at the start of the channel beneath
the input magnet. These spins then diffuse through the spin channel to the
output end and switch the output magnet using spin torque transfer (STT)
mechanism. The structure of a buffer is similar to that of an inverter, but it
uses an applied voltage of $-V_{dd}$ at the input end instead.

\begin{figure}[ht]
\centering
\includegraphics[width=8cm]{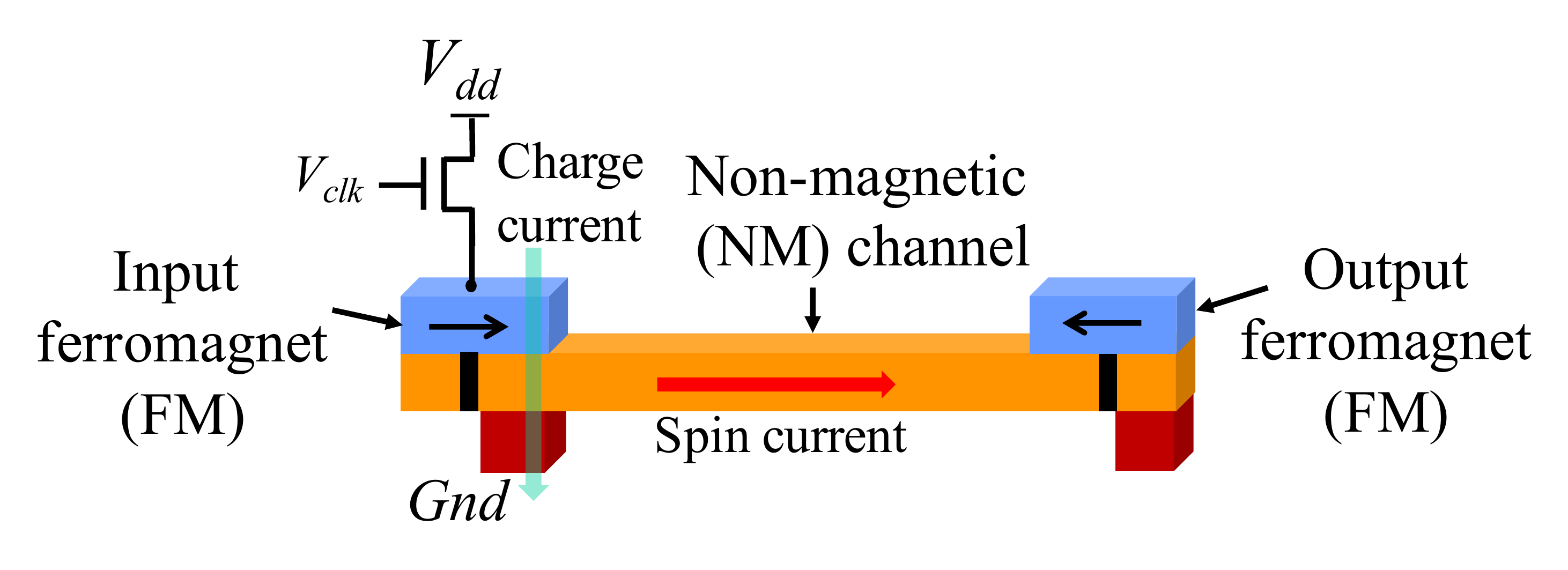}
\caption{A schematic of an all-spin logic inverter.}
\label{fig:ASL}
\end{figure}

In a conventional ASL majority (MAJ) gate~\cite{behin2010proposal}, input
magnets are used to inject spin currents that are sent along a channel. The
spin currents from each of the input magnets compete, and the net spin current
that reaches the output end is an algebraic sum of the spin current from each
input magnet to the output.  The delay is a function of the input vector combination.
Combinations where all inputs assume the same logic value inject noncompeting
spin currents.  For these cases, the net current at the output is larger, and
hence the delay is lower than cases where some spin current contributions
cancel others.
We propose an alternative logic implementation that exploits the timing
difference between the evaluation of the output for various logic states.
We refer to our approach as STEM, a \underline{S}cheme for
\underline{T}wo-phase \underline{E}valuation of \underline{M}ajority logic.
Under this scheme, the output magnet is initialized to a predefined logic stage
in the first phase; in the second phase, currents from a subset of input
magnets compete to evaluate the output to its correct logical value. 

Two-phase schemes have been used in the past in the context of majority logic.
For QCA~\cite{QCAClocking'03, QCABennettClocking}, the first phase pulls the
quantum nanodots from ``null'' state to ``active'' state by altering their
energy profile, while the second phase evaluates the output logic state through
the interaction with neighboring quantum nanodots. Recent work
in~\cite{DominoRoy} uses a two-phase scheme for ASL logic, where a bias voltage
applied to the magnets in the first stage orients the magnetization along the
hard axis, and the output is evaluated in the second stage to switch the output
magnet to the correct state.  In short, all of these approaches initialize the
system to a state between the logic 0 and logic 1 states so that part of the
transition is completed during the first phase.  In contrast, we do not attempt
to place the gate in such an intermediate state, which can sometimes be
unstable.  Furthermore, unlike these prior approaches, our work explicitly
leverages the delay dependence of the logic gate on the input vector to develop
faster and more energy-efficient logic.

The remainder of the paper is organized as follows.  In
Section~\ref{sec:majority}, we describe the conventional implementation of
majority logic in ASL and how such an implementation can be used to build
(N)AND/(N)OR gates. Next, an overview of STEM is provided in
Section~\ref{sec:twophase_intro}, followed by a more detailed discussion of its
circuit-level implementation in Section~\ref{sec:circuit}. A detailed
performance model, described in Section~\ref{sec:model}, is then used to apply
STEM on a set of standard logic functions and a full adder in
Section~\ref{sec:results}.  Finally, concluding remarks are presented in
Section~\ref{sec:conclusion}.

\ignore{
We demonstrate our method on a set of standard logic functions.  In comparison
with the conventional implementation, our method yields gates that are
1.6--3.4$\times$ faster and 1.9--6.9$\times$ more energy-efficient.
When applied to a five-magnet implementation of a full adder, our approach is
shown to be 1.6$\times$ faster and 2.8$\times$ more energy-efficient.

\begin{itemize}
\item
Majority logic widely used (quantum nanodots, NML, ASL, CSL, ME logic)
\item
Speed and power are the major limitations: desirable to compete with CMOS on at least one front
\item
Current approaches wasteful due to the need for evaluating every element
\item
Description of a typical approach
\item
Outline of our approach: two-phase logic
\item
Related work: hard axis initialization in ASL paper, Kaushik's papers (one old, one domino)
\end{itemize}}

\section{Conventional majority logic implementations}
\label{sec:majority}

In this section, we first demonstrate the basics of the majority logic function used
to implement spin-based majority gates or AND/OR/NAND/NOR gates, and then show
that this structure can result in large disparities in the delay, depending on
the input vector combination. Due to this disparity, in many cases the circuit may be
active for an unnecessarily long time, sinking power.

\subsection{The majority logic function}
\label{sec:majorityA}

The majority logic function, MAJ$k$ operates on a set of $k$ binary logic
inputs and outputs the value that represents the majority. For an odd number of
inputs, the output is always binary.  A three-input majority logic function,
MAJ3(A,B,C), operates on Boolean inputs A, B, and C to produce an output Z, as
illustrated in Fig.~\ref{fig:MAJ_conventional_fig}. The corresponding truth
table in Fig.~\ref{fig:MAJ_conventional_tt} illustrates the notion that the
signal that propagates to the output is 3$\times$ stronger when all inputs are
identical as compared to the other cases, and to a first order, this translates
to a 3$\times$ faster switching speed for the stronger signal.

\setcounter{figure}{2}
\begin{figure}[ht]
\centering
\subfigure[]{
\includegraphics[width=3.5cm]{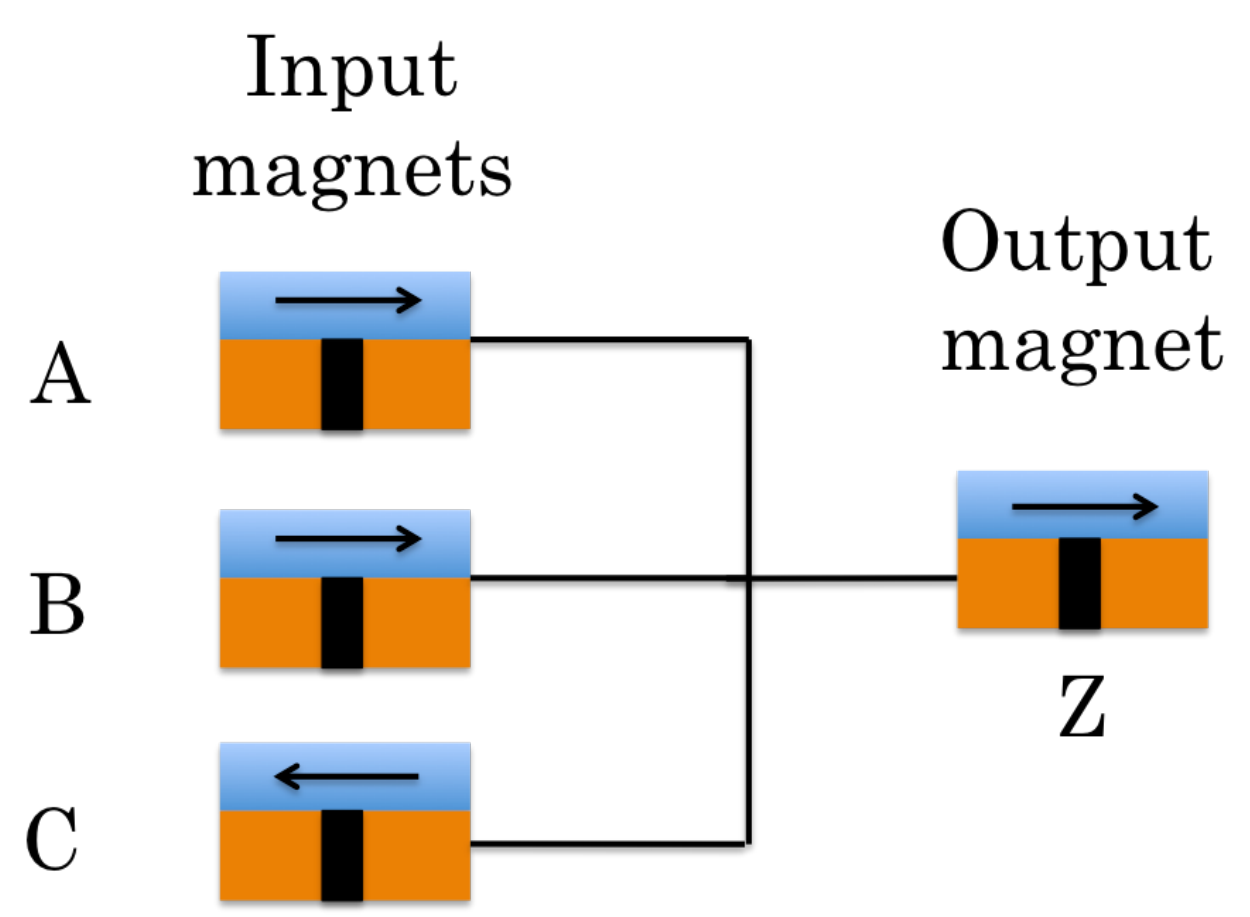}
\label{fig:MAJ_conventional_fig}
}
\subfigure[]{
\scriptsize
\begin{tabular}{||ccc|c||c||}
\hline
A & B & C & Z & Strength  \\
  &   &   &   & $\propto$ 1/Delay  \\
\hline
0 & 0 & 0 & 0 & 3$\times$ \\
0 & 0 & 1 & 0 & 1$\times$ \\
0 & 1 & 0 & 0 & 1$\times$ \\
0 & 1 & 1 & 1 & 1$\times$ \\
1 & 0 & 0 & 0 & 1$\times$ \\
1 & 0 & 1 & 1 & 1$\times$ \\
1 & 1 & 0 & 1 & 1$\times$ \\
1 & 1 & 1 & 1 & 3$\times$ \\
\hline
\end{tabular}
\label{fig:MAJ_conventional_tt}
}
\setcounter{figure}{1}
\caption{(a) The MAJ3(A,B,C) function and (b) its truth table, indicating the
delay associated with evaluating each state.}
\end{figure}

\setcounter{figure}{3}
\begin{figure}[ht]
\centering
\subfigure[]{
\includegraphics[width=4.1cm]{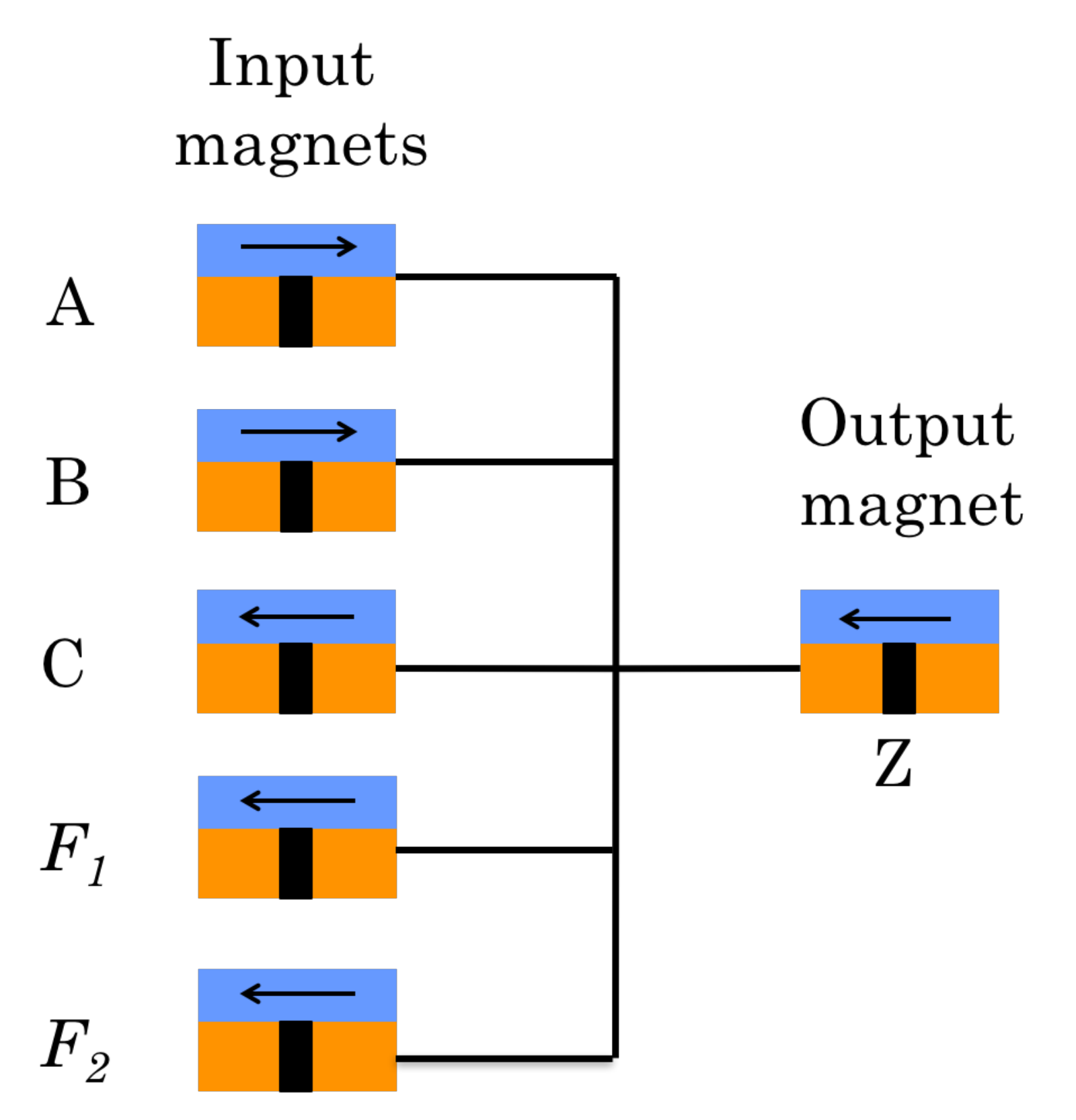}
\label{fig:AND3_conventional_fig}
}
\subfigure[]{
\scriptsize
\centering
\begin{tabular}{||ccc|cc||c|c||}
\hline
A & B & C & $F_1$ & $F_2$ & Z & Strength  \\
  &   &   &       &       &   & $\propto$ 1/Delay  \\
\hline
0 & 0 & 0 & 0     &  0    & 0 & 5$\times$ \\
0 & 0 & 1 & 0     &  0    & 0 & 3$\times$ \\
0 & 1 & 0 & 0     &  0    & 0 & 3$\times$ \\
0 & 1 & 1 & 0     &  0    & 0 & 1$\times$ \\
1 & 0 & 0 & 0     &  0    & 0 & 3$\times$ \\
1 & 0 & 1 & 0     &  0    & 0 & 1$\times$ \\
1 & 1 & 0 & 0     &  0    & 0 & 1$\times$ \\
1 & 1 & 1 & 0     &  0    & 1 & 1$\times$ \\
\hline
\end{tabular}
\label{fig:AND3_conventional_tt}
}
\setcounter{figure}{2}
\caption{(a) The AND3 function A$\wedge$B$\wedge$C implemented with a MAJ5 gate,
and (b) the analysis of its delay as a function of the input state.}
\end{figure}

\subsection{Realizing AND/OR gates using majority logic}
\label{sec:majorityB}

In prior work, techniques for implementing $n$-input gates that realize AND,
OR, NAND, and NOR functionalities based on majority logic primitives have
been proposed \cite{Kim2015}.  An $n$-input AND gate, illustrated in
Fig.~\ref{fig:AND3_conventional_fig} for $n=3$, is realized using a
majority gate with $(2n-1)$ inputs, which include the $n$ inputs of the AND
gate and $(n-1)$ fixed inputs at logic 0. The majority function achieves a
value of 0 only when all $n$ inputs to the AND gate are at logic 1.  Similarly,
an $n$-input OR gate augments the $n$ inputs with $(n-1)$ fixed inputs at logic
1.  The implementation of NAND and NOR gates requires an inversion after the
AND and OR functionalities, respectively, and in some technologies, this
inversion can be applied inexpensively. For instance, as indicated in
Section~\ref{sec:intro}, an ASL implementation simply requires an inversion of
the $V_{dd}$ and $V_{ss}$ polarities for the gate.

Fig.~\ref{fig:AND3_conventional_tt} illustrates the net signal strength
associated with the majority function evaluation for an AND3 gate using a MAJ5
structure. Since the strength of the signal can vary by 5$\times$ over all inputs,
this implies that the gate delay can be 5$\times$ faster for the 000 case, as
compared to the 011, 101, 110, and 111 inputs.

\section{STEM: A two-phase majority logic scheme}
\label{sec:twophase_intro}

In this section, we will describe an alternative implementation of the gates
described above. The method proceeds in two phases:
\begin{itemize}
\item
{\em Phase 1: Initialization}: The output is initialized to a specific value.
\item
{\em Phase 2: Evaluation}: The gate inputs are applied to potentially update the output,
but the evaluation step is terminated at time $T_{eval}$. 
\end{itemize}
This method is superficially similar to the idea of domino logic in CMOS
circuits, where the output is initially precharged to logic 1, and then evaluated and
conditionally discharged. However, in domino logic, the initialization phase
typically corresponds to a precharge that sets the output to logic 1, which is
conditionally discharged during the evaluation phase; as we will see, the approach
used here is different.

\setcounter{figure}{4}
\begin{figure}[ht]
\centering
\subfigure[]{
\includegraphics[width=3.8cm]{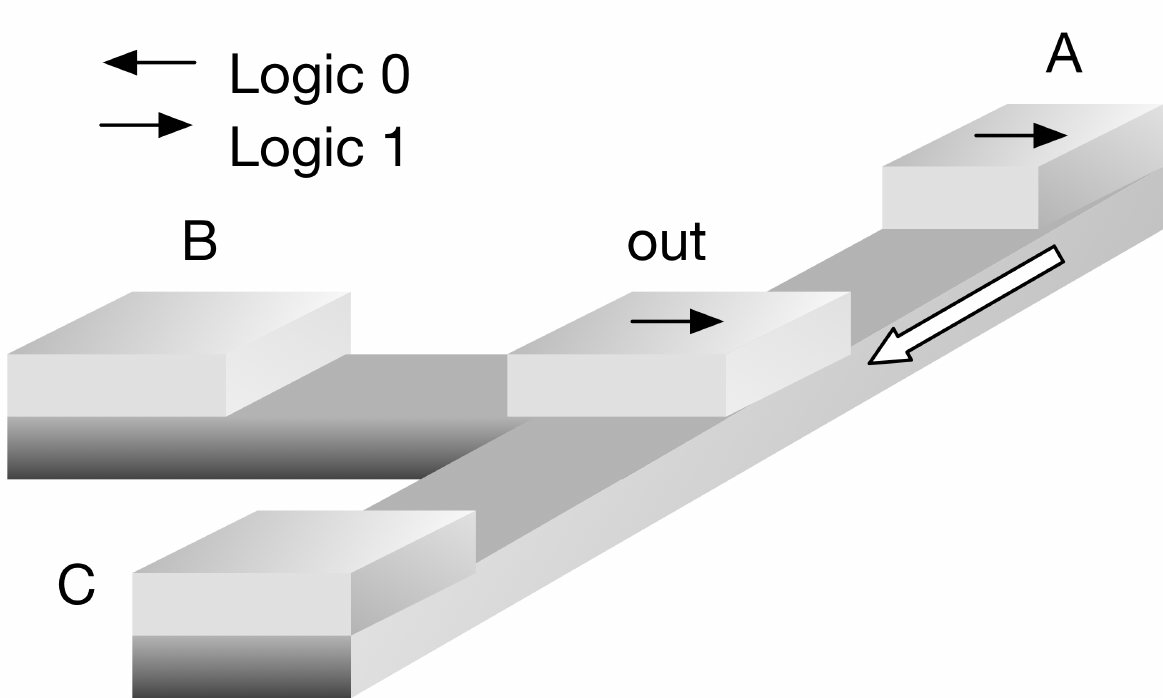}
\label{fig:twophase_init}
}
\subfigure[]{
\includegraphics[width=3.8cm]{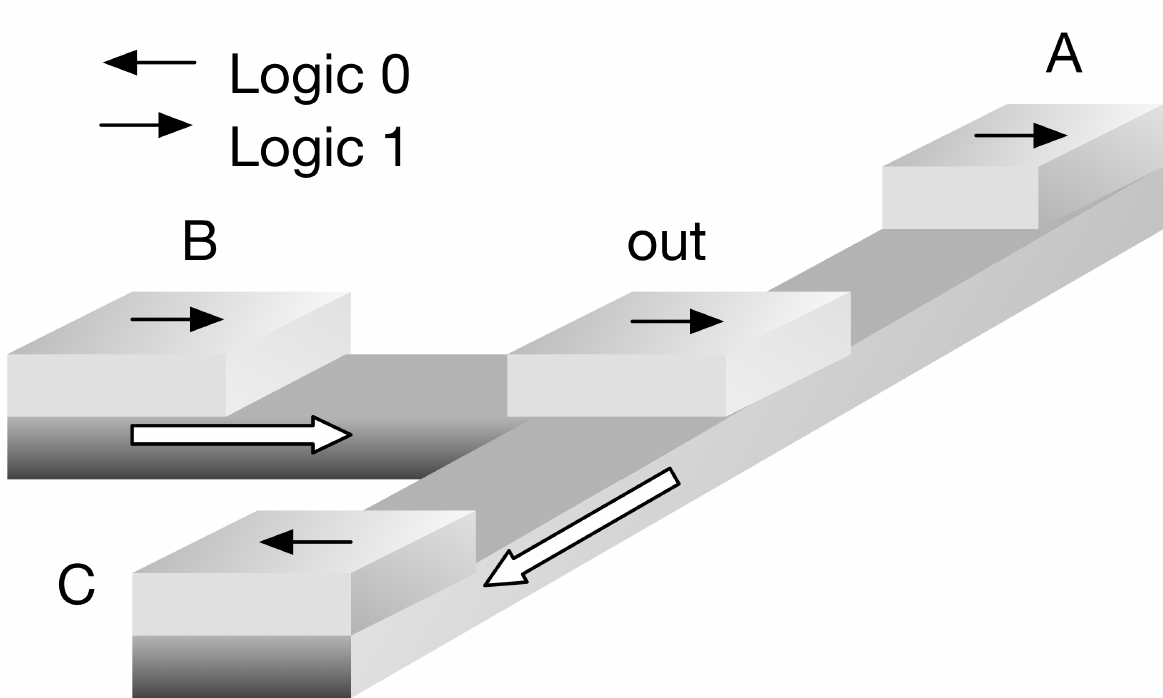}
\label{fig:twophase_eval}
}
\label{fig:MAJ3_twophase}
\setcounter{figure}{3}
\caption{Two-phase operation for the MAJ3 gate under inputs 010: (a) the
initialization phase and (b) the evaluation phase.}
\end{figure}

We demonstrate our idea on an implementation of the majority function,
MAJ3(A,B,C).  In the initialization phase the output is set to the value of one
of the inputs, say A, as shown in Fig.~\ref{fig:twophase_init}.  During the
evaluation phase, the other two inputs are applied to a two-input majority gate
that drives the output, as illustrated in Fig.~\ref{fig:twophase_eval}. If the
two inputs are the same, they clearly form a majority and overwrite the output.
If they are unequal, then they leave the initialization undisturbed, again
resulting in the correct output. Note that the only case in which the logic is
switched is when the inputs B and C both disagree with input A, and the
switching time is $2 T_{eval}$, based on the assumptions in
Fig.~\ref{fig:AND3_conventional_tt}.

In the conventional scheme, the magnets corresponding to A, B, and C must be
symmetric and must contribute an equal spin current. However, in our scheme,
the precharging magnet, corresponding to A above, could be arbitrarily strong
and could potentially perform precharge very quickly. The other magnets,
corresponding to B and C in our example, must be symmetric and could be
smaller, to achieve a better power-delay tradeoff.

\begin{table}[ht]
\centering
\begin{tabular}{||ccc|c||c||}
\hline
A & B & C & Output    & Output    \\
  &   &   & (Phase 1) & (Phase 2) \\
\hline
0 & 0 & 0 & 0         &  0 \\
0 & 0 & 1 & 0         &  0 \\
0 & 1 & 0 & 0         &  0 \\
0 & 1 & 1 & 0         &  1 \\
1 & 0 & 0 & 1         &  0 \\
1 & 0 & 1 & 1         &  1 \\
1 & 1 & 0 & 1         &  1 \\
1 & 1 & 1 & 1         &  1 \\
\hline
\end{tabular}
\caption{The proposed STEM method for implementing the majority function,
MAJ3(A,B,C).}
\label{tbl:MAJ_alt}
\end{table}

An AND2 gate is implemented in majority logic as a MAJ3 gate with one input
set to logic 0: this idea can therefore be used above. In this case, the constant
logic 0 input can be used in the precharge phase, so that precharge could
proceed in parallel with computing the values of the inputs to the AND2 gate,
which may come from a prior logic stage.

For an AND gate with a larger number of inputs, such as an AND3 gate, a similar
idea may be used, but now an evaluation deadline is introduced.  For the AND3 gate,
for all eight input combinations for such a gate, the results at the end of
Phases 1 and 2 are shown in Table~\ref{tbl:AND3_alt}.  Note that while several
combinations (011, 101, and 110) would normally evaluate to a majority value of
logic 1, the net spin current, shown in the table, is $3\times$ weaker than the 111
case, implying that the amount of time required to change the state of the output
magnet is about $3 T_{eval}$.  This large gap not only provides fast
evaluation, but also provides a safe margin beyond $T_{eval}$, so that it is
unlikely that marginally early evaluations of the 011, 101, or 110 case (e.g.,
due to process variations) will corrupt the output value.

\begin{figure}[ht]
\centering
\includegraphics[width=5.5cm]{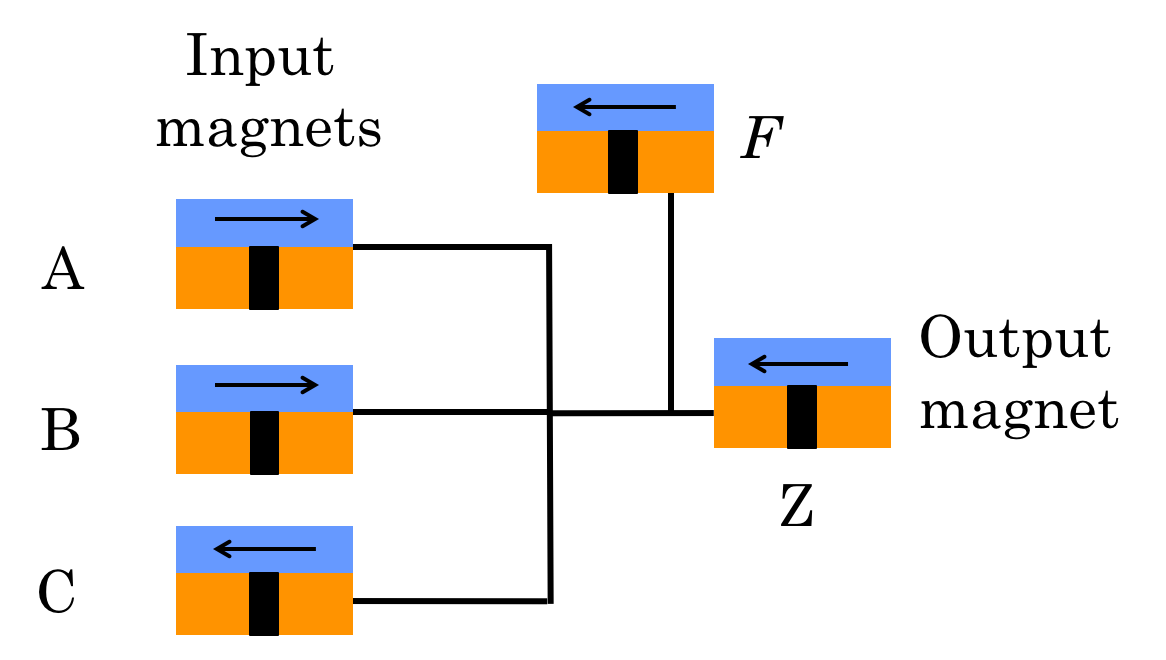}
\caption{A STEM-based implementation of the AND3 gate.}
\label{fig:AND3_twophase}
\end{figure}

The advantage of using the STEM technique is threefold:
\begin{enumerate}
\item
As pointed out above, STEM can improve gate delays.
\item
STEM often requires fewer magnets to implement various gate structures, implying that
the gate area is reduced. For example, for the $n$-input AND gate, a
single fixed input magnet (Fig.~\ref{fig:AND3_twophase}) can initialize the
output magnet to logic state $0$ in Phase 1, as against the $(n-1)$ fixed
inputs required for the conventional ASL implementation, resulting in a savings
of $(n-2)$ magnets.
\item
For AND gates implemented using STEM, this reduction in the number of input
magnets reduces the number of spin currents that are sent to compete at the
output, and thus directly results in a more energy-efficient implementation.
\end{enumerate}

\begin{table}[ht]
\centering
\begin{tabular}{||ccc|c||c|c||}
\hline
A & B & C & Output    & Net spin & Output    \\
  &   &   & (Phase 1) & current  & (Phase 2) \\
\hline
0 & 0 & 0 & 0         &  $-3I_s$ & 0 \\
0 & 0 & 1 & 0         &   $-I_s$ & 0 \\
0 & 1 & 0 & 0         &   $-I_s$ & 0 \\
0 & 1 & 1 & 0         &   $+I_s$ & 0 \\
1 & 0 & 0 & 0         &   $-I_s$ & 0 \\
1 & 0 & 1 & 0         &   $+I_s$ & 0 \\
1 & 1 & 0 & 0         &   $+I_s$ & 0 \\
1 & 1 & 1 & 0         &  $+3I_s$ & 1 \\
\hline
\end{tabular}
\caption{Implementing the AND3 function, A$\wedge$B$\wedge$C, using STEM. Here,
$I_s$ represents the current arriving at the output from a single input
magnet.}
\label{tbl:AND3_alt}
\end{table}

\ignore{
\begin{table}[ht]
\centering
\begin{tabular}{||ccc|cc||c|c|r||}
\hline
A & B & C & $F_1$ & $F_2$ & Output & Net        & \multicolumn{1}{c||}{Delay} \\
  &   &   &       &       & logic  & signal     & \multicolumn{1}{c||}{estimate} \\
\hline
0 & 0 & 0 & 0     &  0    & 0       & 5$\times$ & $T$ \\
0 & 0 & 1 & 0     &  0    & 0       & 3$\times$ & $5/3$ $T$ \\
0 & 1 & 0 & 0     &  0    & 0       & 3$\times$ & $5/3$ $T$ \\
0 & 1 & 1 & 0     &  0    & 0       & 1$\times$ & $5$ $T$ \\
1 & 0 & 0 & 0     &  0    & 0       & 3$\times$ & $5/3$ $T$ \\
1 & 0 & 1 & 0     &  0    & 0       & 1$\times$ & $5$ $T$ \\
1 & 1 & 0 & 0     &  0    & 0       & 1$\times$ & $5$ $T$ \\
1 & 1 & 1 & 0     &  0    & 1       & 1$\times$ & $5$ $T$ \\
\hline
\end{tabular}
\caption{A first-order estimate of delay variability in a majority-gate-based implementation
of an AND3 function, A$\wedge$B$\wedge$C, with two-fixed inputs, $F_1$ and $F_2$, each at logic 0.} 
\label{tbl:AND3}
\end{table}

A significant problem with such majority-based implementations relates to the
large degree of variability associated with the output transitions. 

In effect, the delay corresponds to a single input switching the output.
}

\section{Applying this scheme at the circuit level}
\label{sec:circuit}

\subsection{Gate-level control signals}

The previous section presented an outline of the two-phase STEM scheme. We now
concretely show the use of clocking signals at the gate level that enable
the deployment of this scheme, ensuring that the initialization and evaluation
phases are correctly scheduled.

\begin{figure}[ht]
\centering
\includegraphics[width=8.5cm]{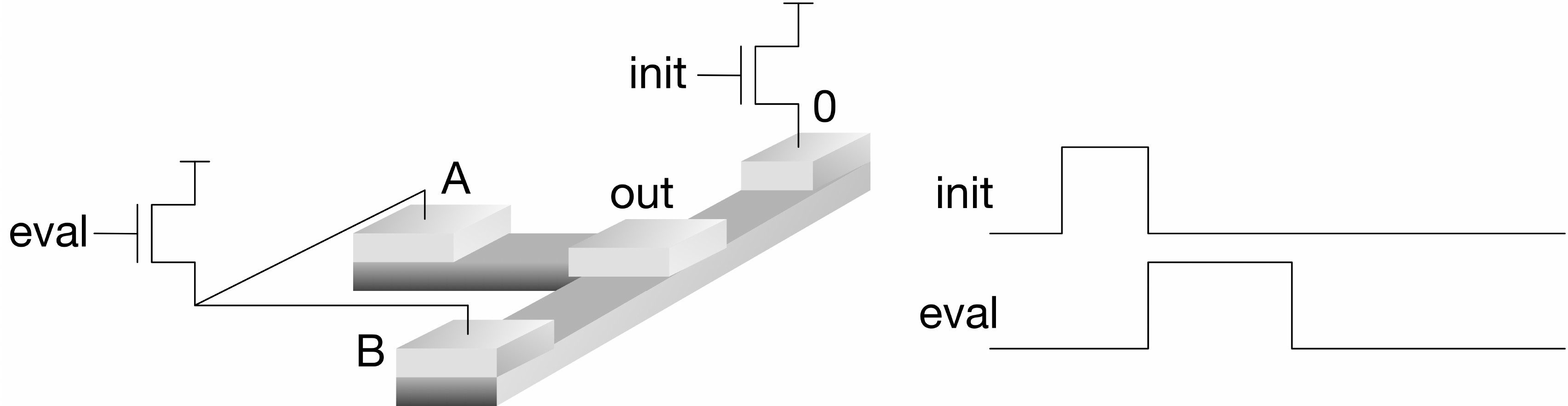}
\caption{A detailed view of the AND2 gate showing control signals for the
initialization and evaluation phases.}
\label{fig:AND2_clocking}
\end{figure}

We show the example of an AND2 gate in Fig.~\ref{fig:AND2_clocking}.
First, the 
``init'' signal is activated during the initialization phase to transmit
data from the fixed-zero input to the output magnet. After
initialization is completed,
the evaluation phase, enabled by the ``eval'' signal, is activated to alter the
initialized output value, if necessary.  Note that for an AND2 gate, the length
of the ``eval'' pulse is unconstrained, but for other types of gates, such as
the AND3 gate described in Table~\ref{tbl:AND3_alt}, the minimum and maximum
pulse width are both constrained: under the coarse delay model proposed
earlier, the minimum width of the signal is $T_{eval}$, and the maximum width
is $3 T_{eval}$.

\subsection{A simple example}
\label{sec:timing_example}

\begin{figure}[ht]
\centering
\includegraphics[width=6.5cm]{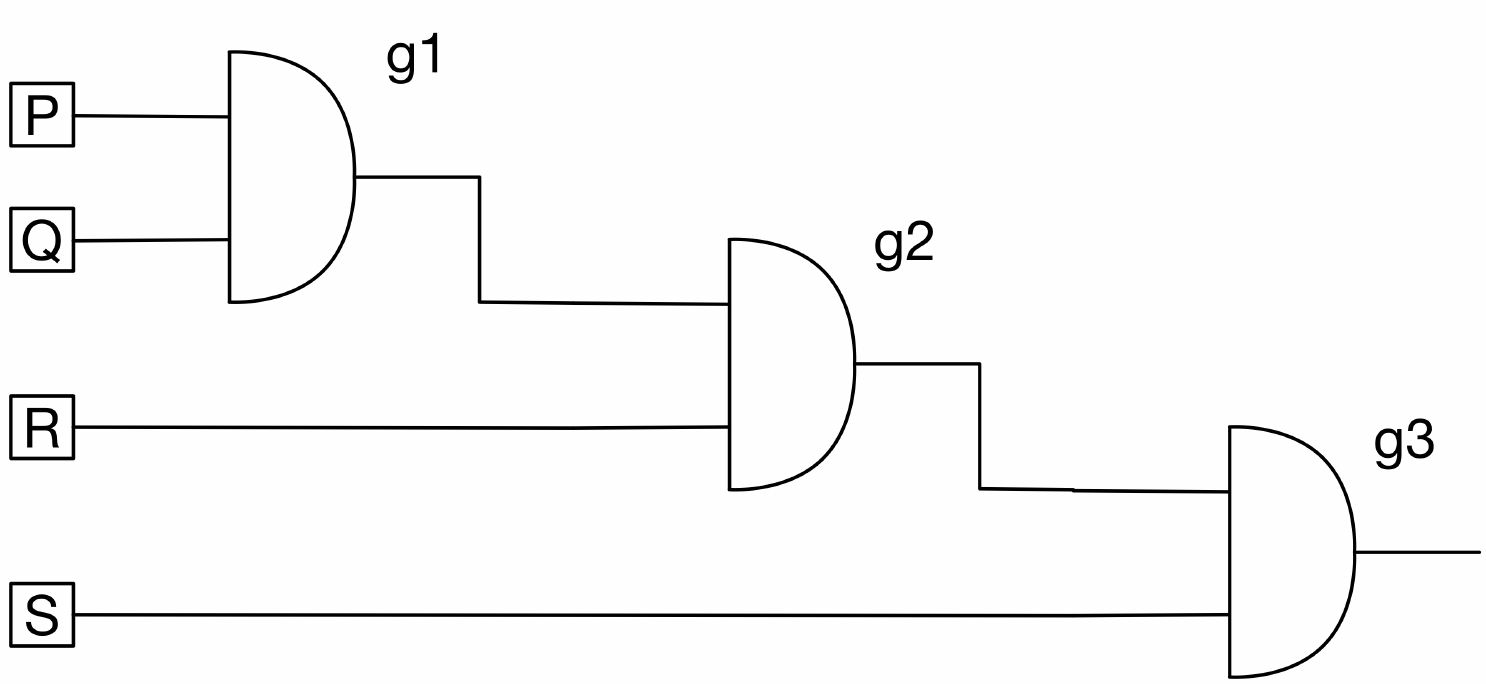}
\caption{An example circuit, implemented using STEM.}
\label{fig:exampleckt1}
\end{figure}

Consider the circuit shown in Fig.~\ref{fig:exampleckt1}, consisting of three
AND2 gates.  The evaluation of the three gates proceeds along the schedule
shown in Fig.~\ref{tbl:AND2computation}, as follows:
\begin{itemize}
\item We assume that the input
magnets connected to the primary inputs P, Q, R, and S are 
initialized when the computation starts. 
\item When the computation begins, 
the gate at the first level, g1, is initialized.
\item After initialization is complete, the gate is evaluated, and simultaneously,
gate g2 is initialized. 
\item In the next time step, g2 is evaluated and g3 is
initialized, and finally, g3 is evaluated. 
\end{itemize}
Note that when the next level of logic is a (N)AND or (N)OR gate, its
initialized value is a constant. Therefore, for such scenarios, it is possible
to simultaneously initialize the next level of logic in parallel with
the evaluation of the current level. For a majority gate,
such simultaneous initialization is not possible, as we will see in the next
subsection.

The timing of these signals is detailed in Fig.~\ref{fig:exampleckt1-timing}.
We denote the widths of the initialization and evaluation pulses for each gate
by $t_{init}$ and $t_{eval}$, respectively.  For this simple example, all gates
are identical and therefore have the same value of $t_{init}$ and the same
value of $t_{eval}$; however, in general, $t_{init} \not = t_{eval}$.
The clocking complexity for the ``init'' and ``eval'' signals is
comparable to that of CMOS domino logic, and similar methods can be used for
generating these clock signals globally and distributing them.

Clearly, the time required to evaluate each gate is $t_{init} +
t_{eval}$. Since $t_{init}$ for g2 overlaps with $t_{eval}$ for g1, the
initialization of g2 can occur immediately after g1 is evaluated.
Similarly, the evaluation pulse to g3 is applied $t_{eval}$ units after
the evaluate pulse to g2 is applied. The same logic can be applied to say that the
evaluation pulses to successive stages must also be delayed by time $t_{eval}$.

\setcounter{figure}{8}
\begin{figure}[ht]
\centering
\subfigure[]{
\scriptsize
\begin{tabular}{||c|c|c|c||}
\hline
time  & g1 & g2 & g3 \\
\hline
0     & init &   -- &   -- \\ \hline
$t_1$ & eval & init &   -- \\ \hline
$t_2$ &   -- & eval & init \\ \hline
$t_3$ &   -- &   -- & eval \\ \hline
\end{tabular}
\label{tbl:AND2computation}
}
\subfigure[]{
\includegraphics[width=6.5cm]{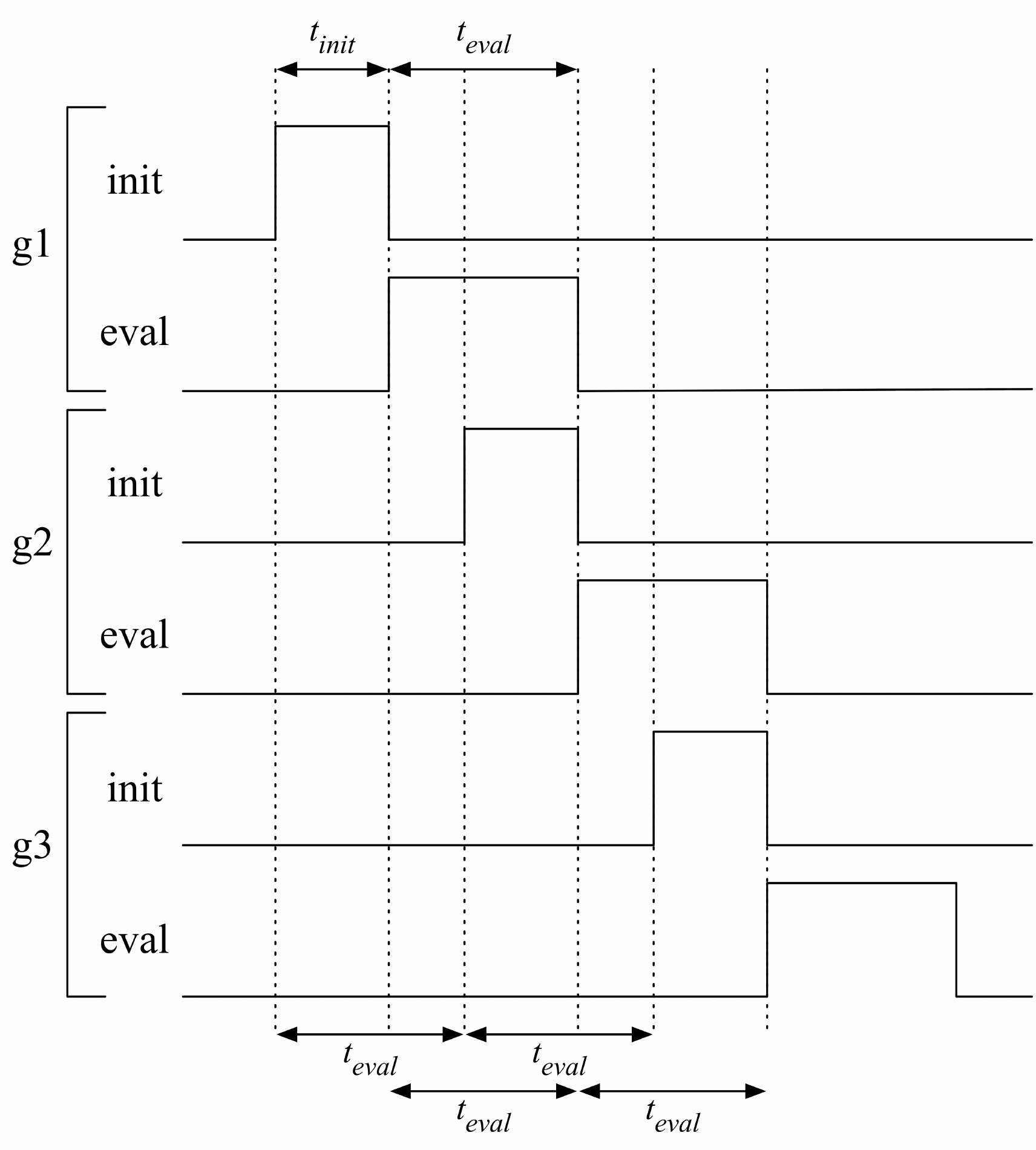}
\label{fig:exampleckt1-timing}
}
\setcounter{figure}{7}
\caption{(a) An overview of the scheduling of the initialization and evaluation
phases for the circuit in Fig.~\ref{fig:exampleckt1}. (b) A timing diagram
showing the ``init'' and ``eval'' signals for each gate.}
\end{figure}

\ignore{
The above sequence determines the scheduling requirements for the ``init'' and
``eval'' signals at each gate.  Based on the above observations, we arrive at
the hardware implementation of the scheme.  The schedule can be implemented
using the scheme in Fig.~\ref{fig:exampleckt1} simply by using delay lines to
transmit the ``init'' and ``eval'' signals from stage to stage, with the delay
lines set to be at least $t_{eval}$.
}

Further, it can be seen that the total evaluation time for this structure (or any
structure with only (N)AND/(N)OR gates) is $t_{init} + 3 t_{eval}$, implying
that the initialization cost must be paid just once for the first gate, and the
evaluation cost dominates for long chains of gates. Note also as observed in
Section~\ref{sec:twophase_intro}, the initialization can generally be made much
faster than evaluation. This can be done since the size of the initializing
magnet can be made much larger than the others and can also be placed closer to
the output magnet as there is no need for symmetry with the other magnets, as
in the conventional implementation.

\subsection{The five-magnet adder}
\label{sec:adder}

\begin{figure}[ht]
\centering
\includegraphics[width=5cm]{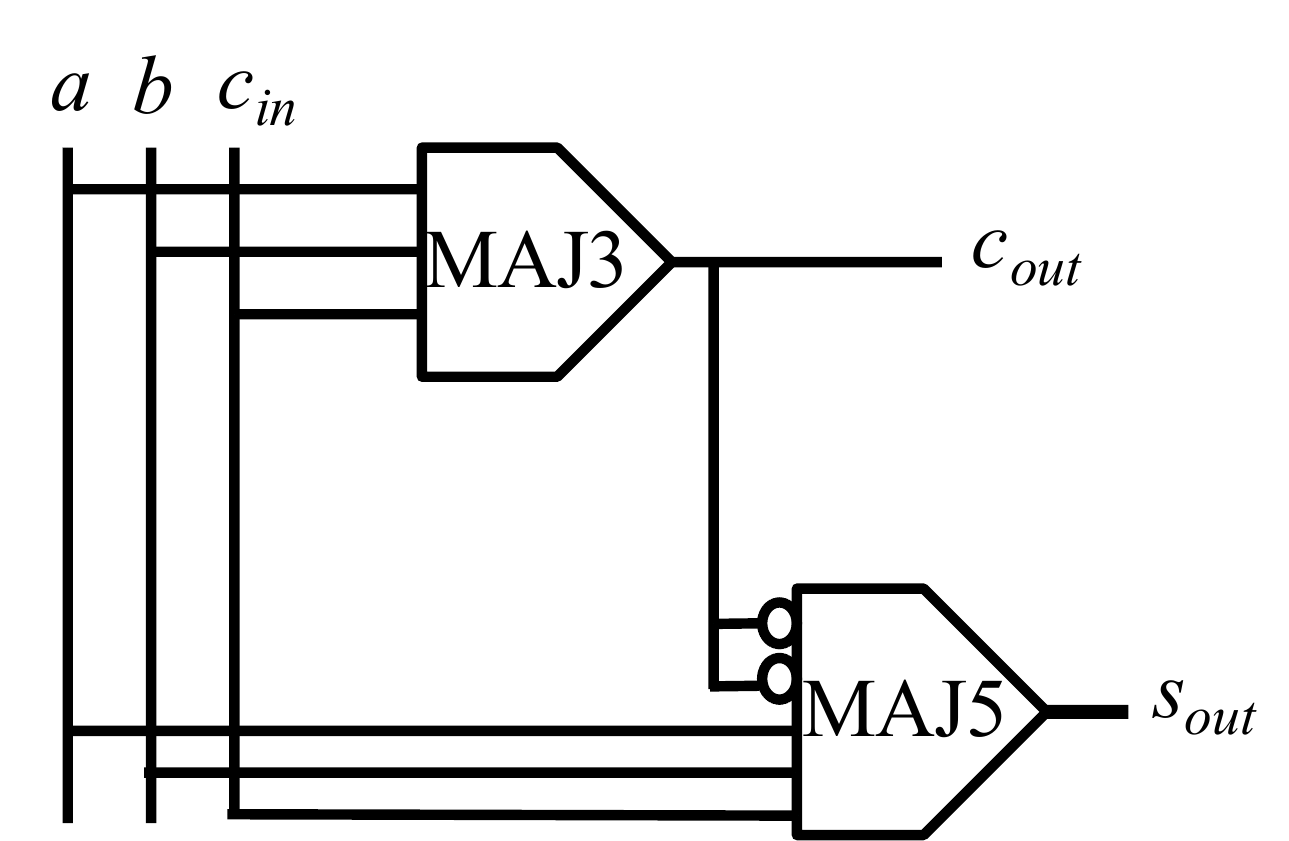}
\caption{The structure of the five-magnet full adder.}
\label{fig:adder}
\end{figure}

The implementation of majority logic allows a spintronic full adder
to be implemented using five magnets~\cite{Martin71,Augustine11}, as shown in 
Fig.~\ref{fig:adder}. We will refer to the corresponding ASL circuit as the
conventional implementation.  In this section, we show that we can achieve a
faster full adder implementation with STEM using the same number of magnets.

\begin{figure}[ht]
\centering
\subfigure[]{
\scriptsize
\begin{tabular}{||ccc||c|c||c||c||}
\hline
$a$ & $b$ & $c_{in}$ & $c_{out}$ & $c'_{out}$ & Strength  & $s_{out}$ \\
\hline
0   & 0   &        0 &         0 &          1 & 3$\times$ & 0 \\
0   & 0   &        1 &         0 &          1 & 1$\times$ & 1 \\
0   & 1   &        0 &         0 &          1 & 1$\times$ & 1 \\
0   & 1   &        1 &         1 &          0 & 1$\times$ & 0 \\
1   & 0   &        0 &         0 &          1 & 1$\times$ & 1 \\
1   & 0   &        1 &         1 &          0 & 1$\times$ & 0 \\
1   & 1   &        0 &         1 &          0 & 1$\times$ & 0 \\
1   & 1   &        1 &         1 &          0 & 3$\times$ & 1 \\
\hline
\end{tabular}
}
\caption{The truth table for $s_{out}$ in the five-magnet full adder.}
\label{fig:five_magnet}
\end{figure}

The principle of the conventional five-magnet adder is that it operates in two
stages. In the first stage, the value of the output carry, $c_{out}$, is
computed using the function MAJ3($a$,$b$,$c_{in}$).  In the second stage, the
value of the sum, $s_{out}$, is calculated as
MAJ5($a$,$b$,$c_{in}$,$c'_{out}$,$c'_{out}$).  In the conventional
implementation, the delay of the full adder 
is the sum of the delay of MAJ3 gate to calculate $c_{out}$ in the first stage
and the delay of MAJ5 gate to calculate $s_{out}$. Both MAJ3 and MAJ5 gate are
implemented using conventional ASL.  The results of this computation are as
shown in columns 4 and 7 of Fig.~\ref{fig:five_magnet}.  The layout of this
adder is shown in Fig.~\ref{fig:layout_adder}(a).

\setcounter{figure}{11}
\begin{figure}[ht]
\centering
\includegraphics[width=4.0cm]{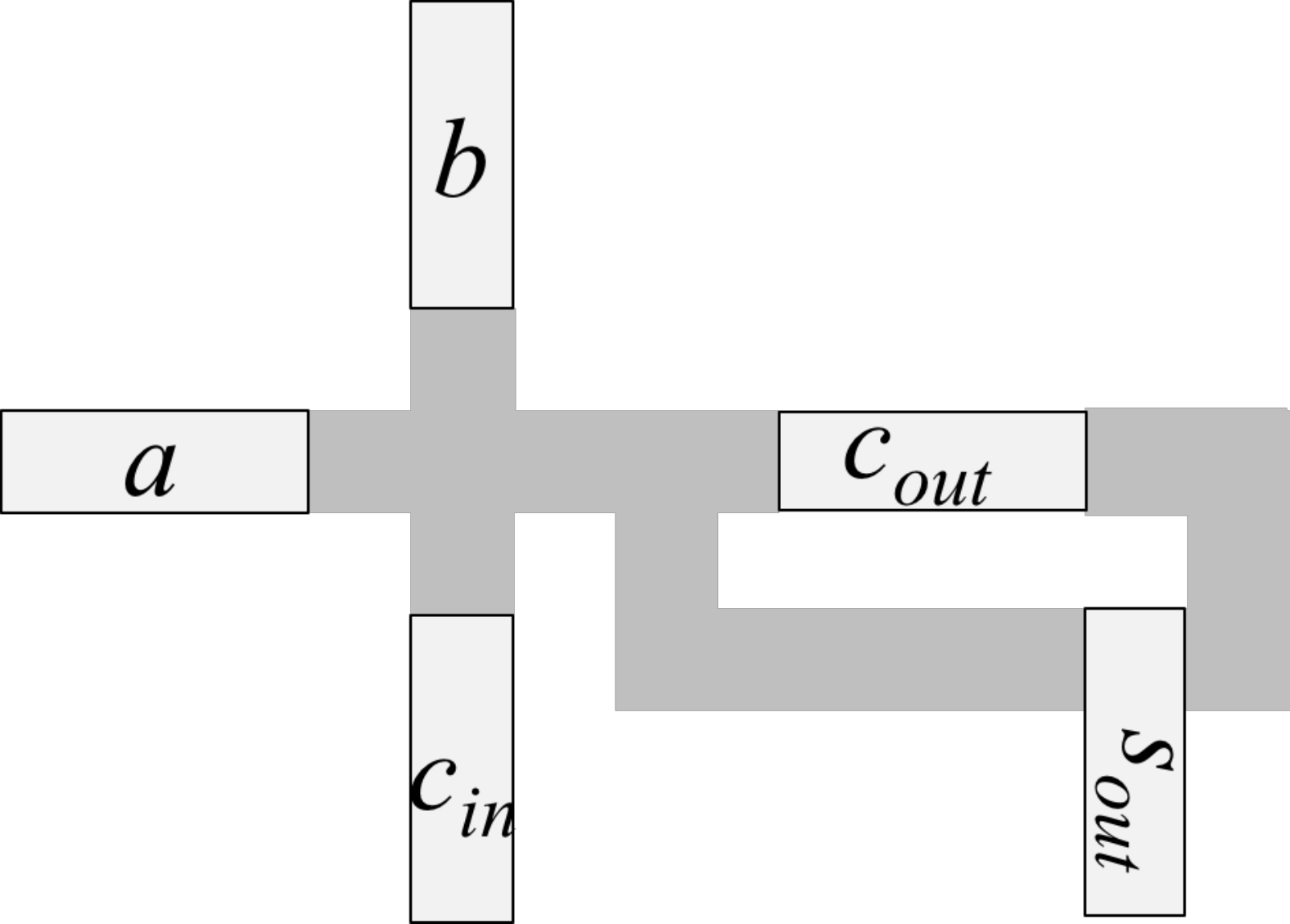}
\hspace{0.5cm}
\includegraphics[width=3.4cm]{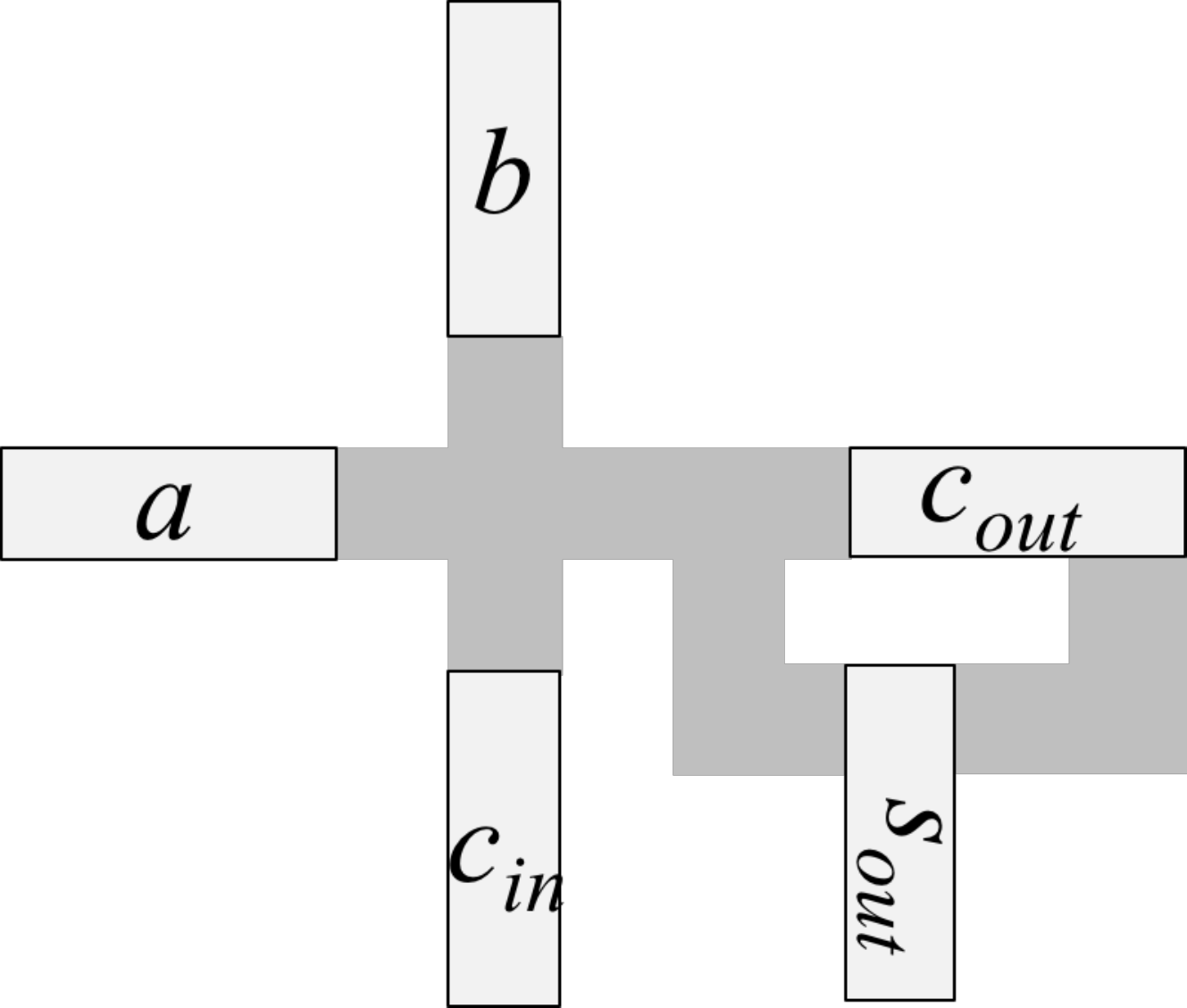}
\\
{\footnotesize (a) \hspace*{3.5cm} (b)}
\setcounter{figure}{10}
\caption{Top view of the layout of the full adder implemented
using (a) the conventional ASL scheme and (b) STEM.}
\label{fig:layout_adder}
\end{figure}

The STEM-based full adder, whose layout is shown in
Fig.~\ref{fig:layout_adder}(b), modifies the above scheme and is implemented as
described below.  As in the conventional adder, the first stage of logic
computes $c_{out}$ as MAJ3($a$,$b$,$c_{in}$), but in this case, the MAJ3 gate
is built using STEM, as explained in Section~\ref{sec:majorityA}.  To implement
the STEM MAJ5 structure, we observe that the original circuit is similar to the
structure in Fig.~\ref{fig:AND3_conventional_fig}, except that the fixed
magnets carry the value of $c_{out}'$. This indicates that a two-phase STEM
structure similar to Fig.~\ref{fig:AND3_twophase}, but with modified
initialization, can be used to implement MAJ5.

The sequence of operations, illustrated in Fig.~\ref{fig:timing_adder1}, is:
\begin{enumerate}
\item
In {\em MAJ3 Phase 1}, input $a$ is first used to initialize $c_{out}$ by applying
an ``init'' signal to magnet $a$.
\item
In {\em MAJ3 Phase 2}, the ``eval'' signal is activated on inputs $b$ and $c_{in}$ 
and their value is used to evaluate and potentially update the $c_{out}$ magnet.
\item
Since the $c_{out}$ signal acts as the initializing signal for $s_{out}$, the
``init'' signal that effects this transfer can only be applied after $c_{out}$
has been computed. Note that this implies that unlike the AND gate example previously
described, there is no overlap with the evaluation phase of the previous stage.
\item Finally, in {\em MAJ5 Phase 2}, we compute MAJ3($a$,$b$,$c_{in}$) and use
the resulting spin current to attempt to update the $s_{out}$ magnet.  Note that
the lengths of the paths from each of these three input magnets to $s_{out}$ are
balanced to ensure equal contributions for the MAJ3 function.
\end{enumerate}
The evaluation times for MAJ3 and MAJ5 are denoted as $t_{eval,1}$ and
$t_{eval,2}$, respectively and will typically be different.  As we will show in
Section~\ref{sec:results_adder}, the timing signals and their sequence can be
optimized to reduce the number of global clock signals, but for this clocking
scheme, let the time required to initialize $c_{out}$ in the
first step and $s_{out}$ in the third step be denoted as $t_{init,1}$
and $t_{init,2}$, respectively. The circuit delay is then given by 
$T_{adder,STEM} = t_{init,1} + t_{init,2} + t_{eval,1} + t_{eval,2}$.

\setcounter{figure}{12}
\begin{figure}[ht]
\centering
\includegraphics[width=2.5in]{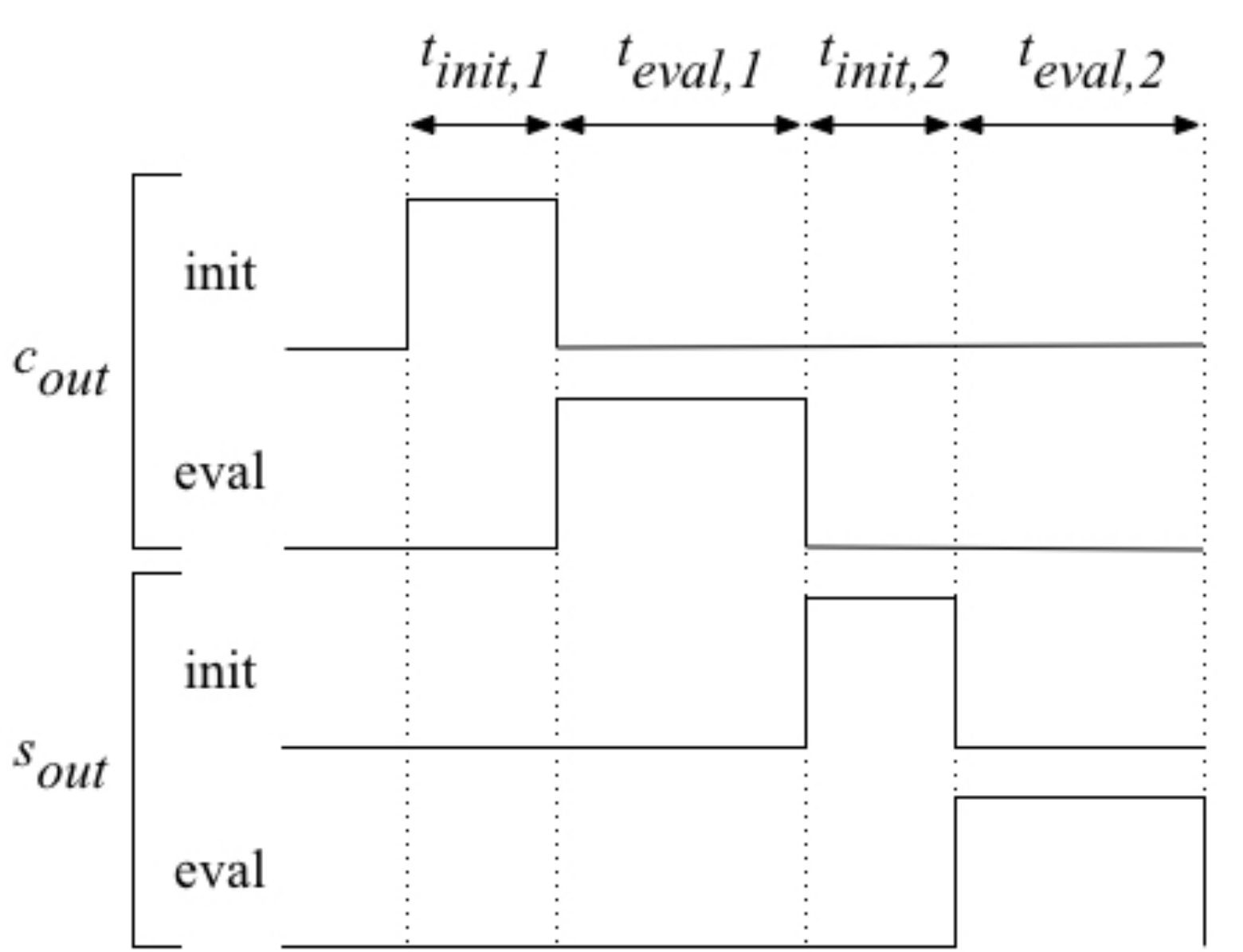}
\setcounter{figure}{11}
\caption{A timing diagram for the five-magnet adder.}
\label{fig:timing_adder1}
\end{figure}

The timing requirements for the fourth step, {\em MAJ5 Phase 2} are similar to those
for the structure in Fig.~\ref{fig:AND3_twophase}, where we allow a limited
time to perform the update. In particular, this time is chosen such that a
3$\times$ current strength (corresponding to the 000 or 111 combinations) can
write the magnet, but any of the other input combinations (which all correspond
to 1$\times$ current strengths) do not have sufficient time to switch the
output magnet.  This places both an upper and lower bound on $t_{eval,2}$, the
time required to implement {\em MAJ5 Phase 2}, but in practice the large gap
between the delay for a 1$\times$ vs. 3$\times$ current implies that the
upper bound is not a significant factor.  For the adder that is evaluated in
Section~\ref{sec:results} using precise timing models, the 1$\times$ signal is
found to be 2.7$\times$ more than $t_{eval,2}$, while the value of $t_{eval,1}$
is similar to that of $t_{eval,2}$.

\ignore{
Since it is preferable to generate a single evaluation pulse for the second and 
third steps above, we set $t_{eval}~=~\max(t_{eval,1},t_{eval,2})$.  The delay
of the full adder is therefore $T_{adder,STEM}  = t_{init}+ 2t_{eval}$. Here,
$t_{init}$ refers to the initialization of the $c_{out}$ magnet of the MAJ3
gate; $t_{eval}$ is the maximum among the times required to evaluate $c_{out}$
and initialize $s_{out}$ or evaluate $s_{out}$, since the same ``eval'' signal
is used by both the logic stages for computation.     
}

\ignore{
A back-of-the-envelope analysis of the speed of this structure recognizes that
the evaluation of each stage is performed about 3$\times$ faster than the
conventional implementation. However, for computation of $s_{out}$, the
output of MAJ3($a$,$b$,$c_{in}$) has to travel an additional distance 
from $c_{out}$ to switch $s_{out}$ in the second stage. Therefore, the
strength of the net spin current that eventually switches $s_{out}$ would be
less than 3$\times$.  The speed of this structure is slightly less than
3$\times$ that of the conventional five-magnet adder, with the reduction
corresponding to the overhead of initializing the first stage; as observed
earlier, this can be made asymmetrically fast.
}

The STEM-based full adder above uses $c_{out}'$ to initialize $s_{out}$.  An
alternative implementation of the full adder using STEM could be to initialize
$s_{out}$ with one of the input magnets, $a$, $b$, or $c_{in}$: let us say we
use $a$.  This implementation would have the advantage of allowing $s_{out}$ to
be initialized in parallel with the evaluation of $c_{out}$ using the MAJ3
gate. However, in the second phase of evaluation, the majority of $b$,
$c_{in}$, $c_{out}'$ and $c_{out}'$ is used to update the $s_{out}$ magnet. The
slowest successful switch corresponds to the case where either $b$ or $c_{in}$
(but not both) agree with $c_{out}'$, resulting in a 3$\times$ switching
current in the direction of $c_{out}'$ and a 1$\times$ current opposing it.
Hence, the net current is 2$\times$, as against the 3$\times$ current for the
implementation above.  This is found to be a dominant factor in determining the
delay of the adder.  Thus, this alternative implementation of the full adder
under STEM is outperformed by the implementation described above.

\section{ASL performance modeling}
\label{sec:model}

\begin{figure}[ht]
\centering
\includegraphics[width=3.5cm]{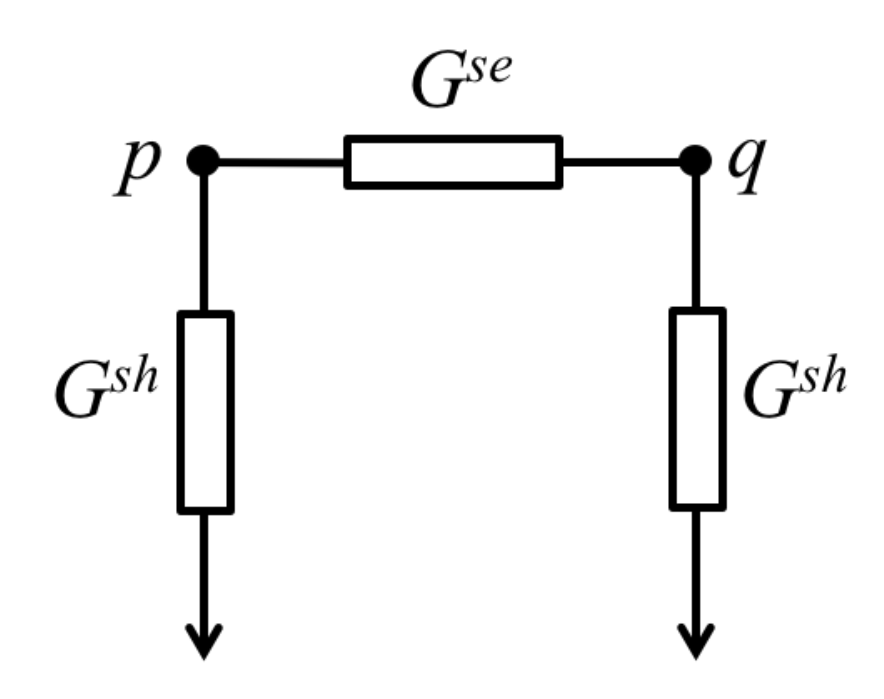}
\caption{$\pi$-model for either FM or NM connected between nodes
{\em p} and {\em q}.}
\label{fig:ckt}
\end{figure}

The metrics used to motivate our approach in the previous sections were
based on coarse estimates based on the spin current.  To evaluate
the approach, it is essential to use accurate models for the delay and
energy of the two-phase and conventional circuit implementations.  In this
section, we provide a brief overview of the spin circuit model used to evaluate
ASL devices.  Based on~\cite{Srinivasan2013}, the ferromagnet (FM) and the
nonmagnetic (NM) channel in Fig.~\ref{fig:ASL} are each represented by a
$\pi$-model, shown in Fig.~\ref{fig:ckt}, where $p$ and $q$ are the end points
of the magnet or the channel in the direction of current flow.  

The series and shunt conductance matrices used in the $\pi$-model for FM and
NM, respectively denoted by $G^{se}$ and $G^{sh}$, are given by:
\begin{align}
G_{FM}^{se}  & =  \frac{A_{FM}}{\rho_{FM} L_{FM}} \begin{bmatrix}
	1 & \beta \\
	\beta &   {\beta}^2 +
		\left ( \frac{(1 - \textit p^2) L_{FM}}{\lambda_{sf,FM}} \right )
			\mathrm{cosech} \left ( \frac{L_{FM}}{\lambda_{sf,FM}} \right )
	\end{bmatrix}
\label{Gmatrixst}
\\
G_{FM}^{sh}  & =  \frac{A_{FM}}{\rho_{FM} L_{FM}} \begin{bmatrix}
	0 & 0 \\
	0 & \left ( \frac{(1- \textit p^2) L_{FM}}{\lambda_{sf,FM}} \right )
		 \tanh \left ( \frac{L_{FM}}{2\lambda_{sf,FM}} \right )
	\end{bmatrix}
\\
G_{NM}^{se}  & =  \frac{A_{NM}}{\rho_{NM} L_{NM}} \begin{bmatrix}
	1 & 0 \\
	0 & \left ( \frac{L_{NM}}{\lambda_{sf,NM}} \right )  \mathrm{cosech}
		 \left ( \frac{L_{NM}}{\lambda_{sf,NM}} \right )
	\end{bmatrix}
\\
G_{NM}^{sh}  & =  \frac{A_{NM}}{\rho_{NM} L_{NM}} \begin{bmatrix}
	0 & 0 \\
	0 & \left ( \frac{L_{NM}}{\lambda_{sf,NM}} \right ) \tanh 
		\left ( \frac{L_{NM}}{2\lambda_{sf,NM}} \right )
	\end{bmatrix}
\label{Gmatrix}
\end{align}

\noindent
where $A$, $\rho$, $\lambda_{sf}$, $p$, and $L$ denote the cross-sectional
area, resistivity, spin-diffusion length, spin polarization, and the length, respectively,
and the subscript denotes whether the attribute corresponds to the FM
or the NM.  The spin current ($I_s$) and the charge current ($I_c$) between
nodes $p$ and $q$ obey the relation:
\begin{align}
\begin{bmatrix}
I_{c,pq} \\
I_{s,pq}
\end{bmatrix}
= &
\begin{bmatrix}
G^{se}
\end{bmatrix}_{2 \times 2}
\begin{bmatrix}
V_{c,p} - V_{c,q} \\
V_{s,p} - V_{s,q}
\end{bmatrix} + 
\begin{bmatrix}
G^{sh}
\end{bmatrix}_{2 \times 2}
\begin{bmatrix}
0 \\
V_{s,p}
\end{bmatrix}
\label{eqn:calcIs}
\end{align}

\noindent
Using these matrices, the stamps for the individual circuit elements are
built to populate the nodal analysis matrix, $G_{ckt}$. The circuit
equation is then given by:
\begin{align}
 \begin{bmatrix}
G_{ckt}
\end{bmatrix}_{2k \times 2k}
{\bf V}
= {\bf I}
\label{eq:Gckt}
\end{align}
where $k$ represents the number of nodes in the circuit, ${\bf V} = [
\vec{V}_1 \cdots \vec{V}_k ]^T$ represents the vector of nodal voltages, 
and $\bf I$ corresponds to the vector of excitations to the circuit. The
system of equations in Equation~\eqref{eq:Gckt} is solved to obtain the
spin and charge nodal voltages and branch currents. The spin current,
$I_s$ at the output magnet is used to calculate the delay of the gate, 
$T_{gate}$ and the switching energy of the gate, $E_{gate}$ as: 
\begin{align}
T_{gate} &= \frac{2qN_sf}{I_s} \label{eq:delay}\\
E_{gate} &= V_{dd}I_cT_{gate} \label{eq:energy}
\end{align} 
\noindent 
where $N_s$ refers to the number of Bohr magnetons in FM, $q$ the
electron charge, $I_c$ the total charge current involved in the
switching process, and $f$ is a constant~\cite{Switching_energy_delay'11}.
We perform the ASL modeling and simulations on MATLAB tool. 

\section{Results}
\label{sec:results}

We now evaluate the delay and energy associated with applying the STEM scheme
in an ASL technology and compare it to the conventional ASL implementation. We
first examine gates that implement basic logic functions and then evaluate the
five-magnet adder.  The simulation parameters for the ASL structures, including
technology-dependent parameters and physical constants, are shown in
Table~\ref{table:Parameters}. 
The input and output FMs are all based on
perpendicular magnetic anisotropy (PMA). In contrast with magnets with in-plane
magnetic anisotropy (IMA) such magnets are more compact since they do not
require a specific aspect ratio in the plan of the layout to achieve shape
anisotropy, and their dipolar coupling to neighboring magnets is weaker~\cite{MegJETC}.

Our evaluations below focus on the energy dissipated in the magnets.
Conventional ASL uses access transistors to control the current sent through
the magnets~\cite{Sharad13}.  To a first order, the energy dissipated in
access/gating transistors is similar in conventional ASL and in STEM since a
similar current is delivered in each case: conventional ASL uses a smaller
number of large access transistors, while STEM uses a larger number of smaller
access transistors. This is due to the fact that in conventional ASL, a single clock
signal clocks all the magnets, while in the case of STEM, a subset of magnets are 
clocked by the ``init'' signal with the rest being clocked by the ``eval'' signal. 
We neglect transistor energy because substantial amounts of
sharing of these transistors is possible in large circuits, and it is difficult
to estimate the energy impact on these smaller circuits. This caveat should be
kept in mind while interpreting the energy numbers reported here, keeping in mind that these
are energy improvements at the gate level. This
assumption does not affect the reported delay improvements.

\begin{table}[hb] 
\centering 
\begin{tabular}{| l | l |}\hline 
\textbf{Parameters } 													& \textbf{Value}\\ \hline \hline
Spin polarization factor, $p$                 							& 0.8\\
\hline
Resistivity of magnet, $\rho_{FM}$ \cite{behin2010proposal}             & $\SI{170}{{\Omega}nm}$ \\ 
\hline
Resistivity of channel, $\rho_{NM}$ \cite{behin2010proposal}            & $\SI{7}{{\Omega}{nm}}$\\
\hline
Spin flip length of magnet, $\lambda_{sf,FM}$ \cite{behin2010proposal}  & $\SI{5}{nm}$\\
\hline
Spin flip length of channel, $\lambda_{sf,NM}$ \cite{behin2010proposal} & $\SI{500}{nm}$\\
\hline
Magnet dimension (length$\times$width$\times$thickness)                 & $30{\times}10{\times}3$ nm$^3$\\
\hline
Channel width                         									& $\SI{10}{nm}$\\
\hline
Channel thickness                     									& $\SI{10}{nm}$\\
\hline
Bohr magneton, $\mu_B$              									& $9.274{\times}10^{-24} JT^{-1}$\\
\hline
Saturation magnetization, $M_s$ \cite{behin2010proposal}     			& $\SI{780}{emu/cc}$\\
\hline
Charge of an electron               									& $\SI{1.6e-19}{C}$\\
\hline
Input voltage                       									& $\SI{100}{mV}$\\
\hline
\end{tabular}
\caption{Parameters used to model ASL structures in our simulations.}
\label{table:Parameters}
\end{table}

\subsection{Evaluating STEM on individual logic gates}

We examine the performance of gates that implement basic logic functionalities, as
described in Section~\ref{sec:twophase_intro}.  We focus on AND and MAJ gates with 
various numbers of inputs. The implementation of NAND, NOR, and OR gates is very 
similar to AND gates, and the results for the AND gate carry over to these types
of gates. We obtain the delay and energy of the AND and MAJ
gates modeled using the method
described in Section~\ref{sec:model} in MATLAB.

For both the AND and MAJ gates, during Phase 1, only one magnet initializes the
output magnet.  Therefore, the initialization delay, $t_{init}$ is equivalent
to one inverter delay.  As stated in Section~\ref{sec:twophase_intro}, unlike
the conventional ASL implementation which requires all input magnets to have the
same size, the initializing input is freed of this constraint since it does not
compete with any other magnet.  Therefore, although
Fig.~\ref{fig:twophase_init} shows the initializing magnet to be unit-sized, in
principle it is possible to upsize this magnet to speed up initialization.  We
consider various scenarios where the strength of the initialization magnet can
be increased to $Q$$\times$ by increasing the area (length $\times$ width) of
the unit-sized magnet (whose dimensions are defined in
Table~\ref{table:Parameters}) by a factor of $Q$, where $Q = [2,4,8,16]$.
Regardless of this choice, in Phase 2, input magnets of size 1$\times$ will
evaluate and switch the output magnet to the correct state.  

\begin{figure}[ht]
    \centering
    \subfigure[]{
	\includegraphics[width=6.5cm]{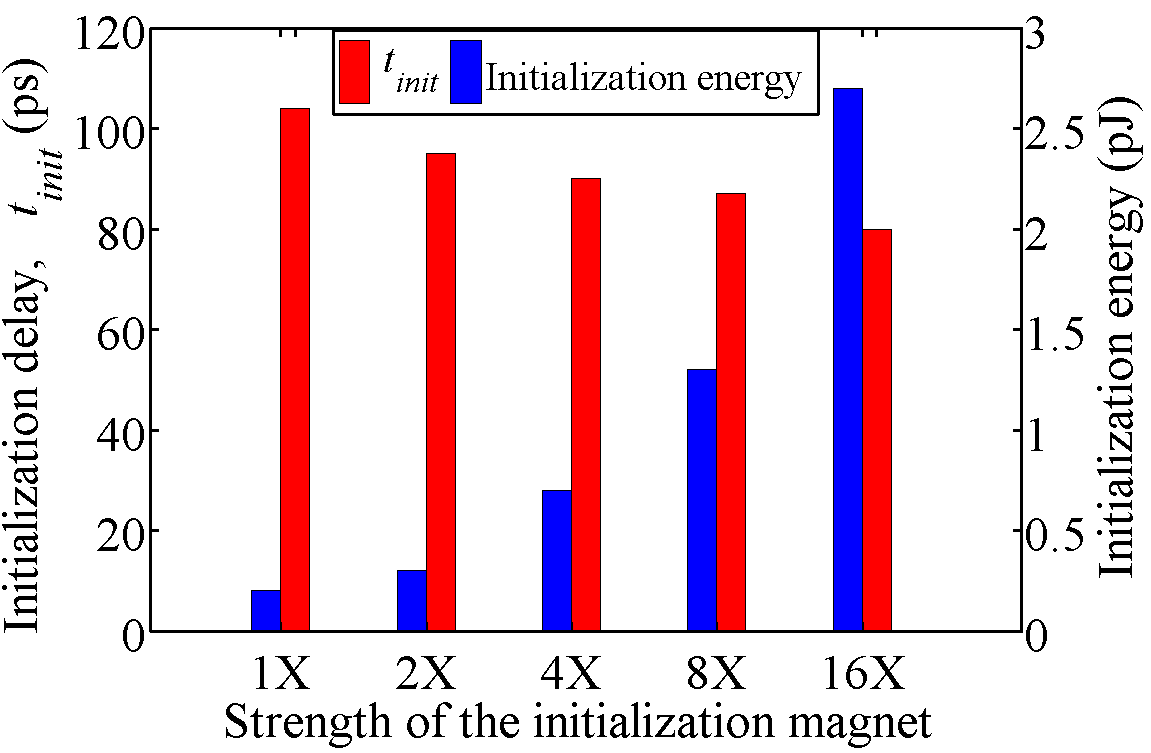}
	}
    \subfigure[]{
	\includegraphics[width=6.5cm]{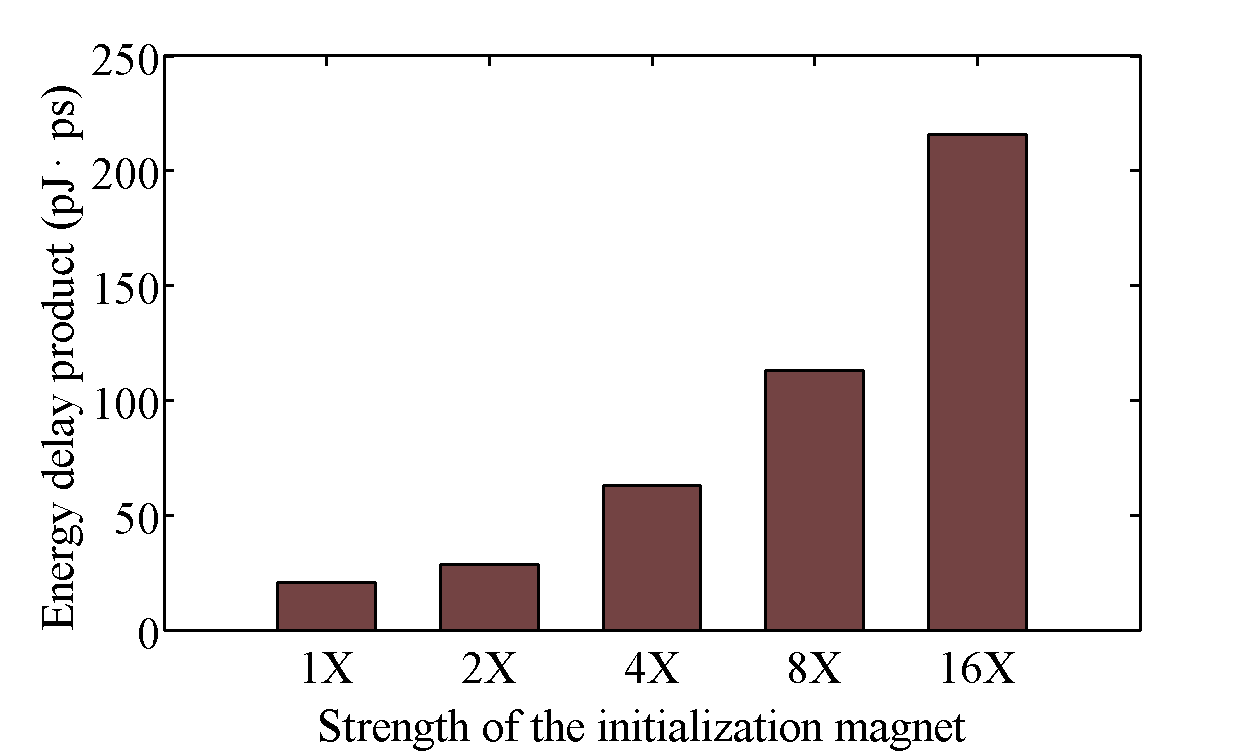}
	}
    \caption{(a) The delay and energy and (b) the energy-delay product for
	the initialization phase, as a function of the strength of the
	initialization magnet.}
    \label{fig:tinit}
\end{figure}

\begin{figure}[ht]
\centering
    \subfigure[] {
        \includegraphics[width=9cm]{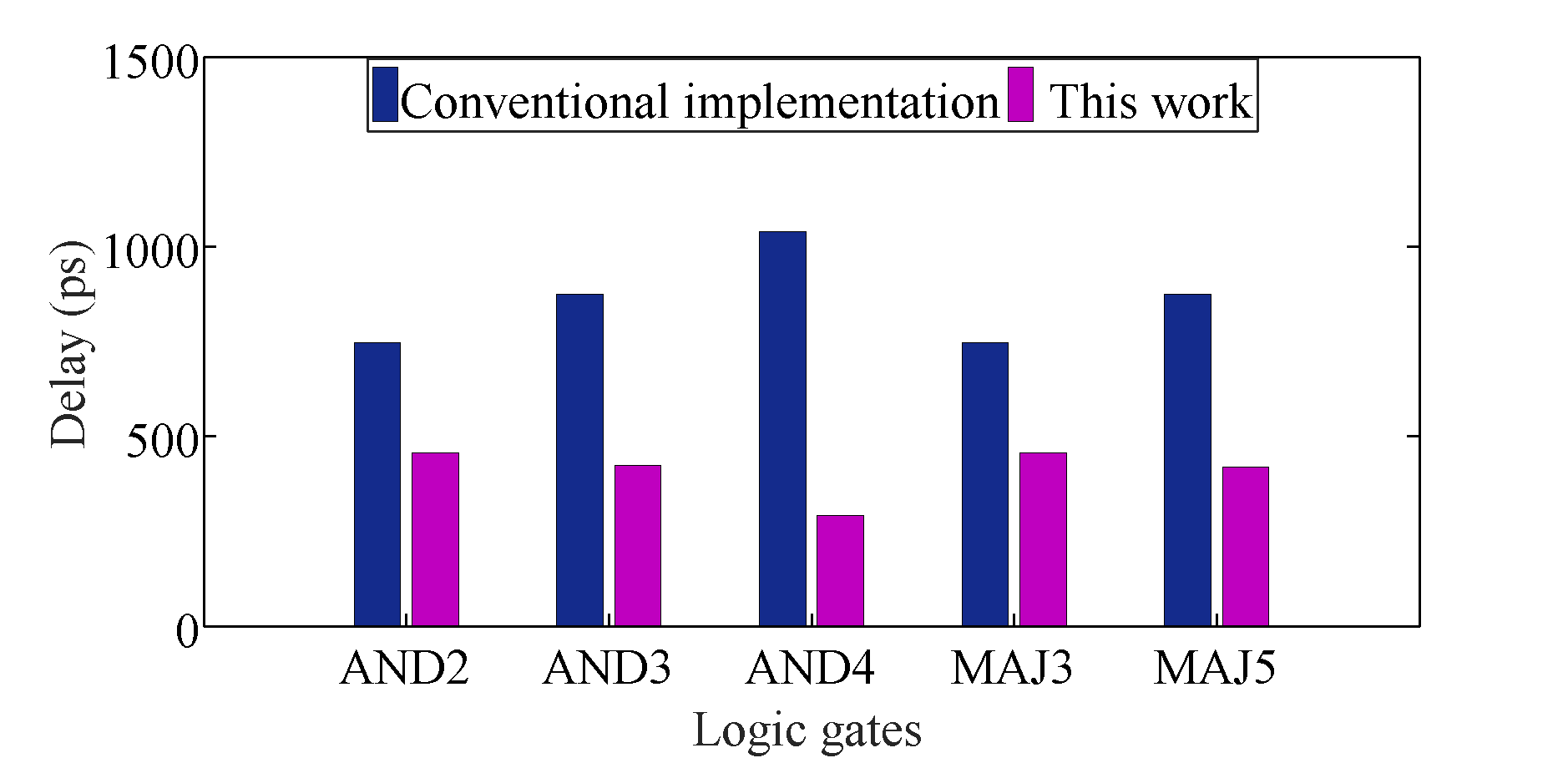}
	}
    \subfigure[] {
        \includegraphics[width=9cm]{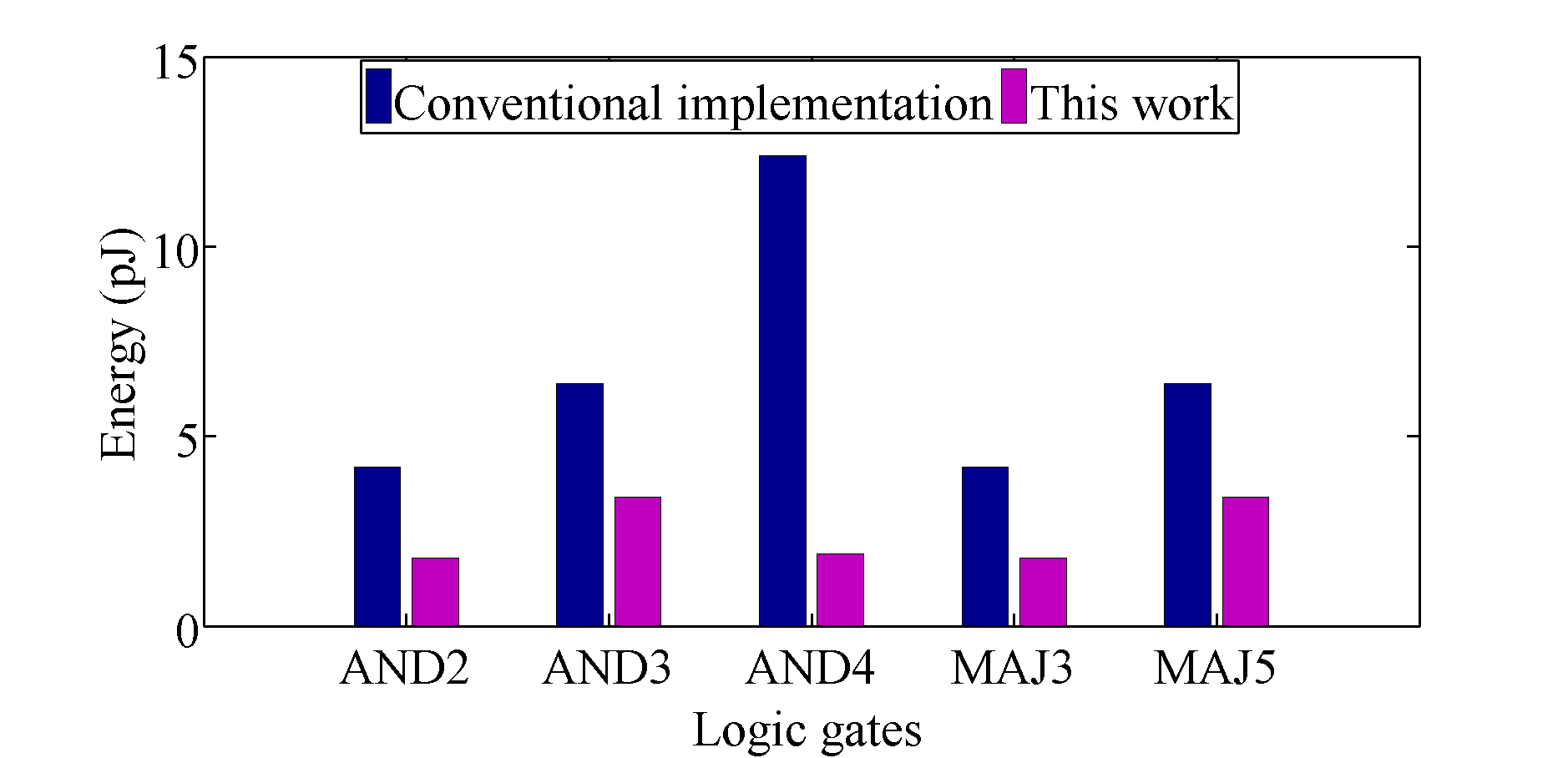}
    
	}
	\caption{Comparison between the (a) delay and (b) energy of the
	conventional ASL implementation and the proposed STEM approach.}
    \label{fig:delay_energy}
\end{figure}

The energy, delay, and energy-delay product for initialization are plotted in
Fig.~\ref{fig:tinit}.  In this work, we choose the solution with the optimal
energy-delay product, corresponding to a 1$\times$ magnet.  This
result can be explained by the fact that due to spin losses in the channel,
increasing the input magnet size does not reduce the delay sufficiently to
compensate for the corresponding increase in energy. Therefore, we use the
1$\times$ magnet for initialization, which corresponds to $t_{init} = 104$ps
and initialization energy of $0.2$pJ.  In other words, based on this analysis,
although it is possible to speed up initialization, we choose to go with the
solution that is similar to the conventional ASL scheme.  The circuit-level
analysis in Section~\ref{sec:timing_example} as well as the adder optimizations
to be proposed in Section~\ref{sec:results_adder} both support this choice,
since the initialization phase can be overlapped with evaluation for all but
the first stage of logic. If it is important to reduce the circuit
delay, the $t_{init}$ value of {\em only} the first logic stage can be made
faster, at the expense of a special narrow clock signal for this stage.

We compare the delay and energy associated with the conventional ASL implementation
and STEM with respect to the AND2, AND3, AND4, MAJ3, and MAJ5 logic gates, and
display the results in Fig.~\ref{fig:delay_energy}. For each logic gate, the
delay for the STEM implementation refers to the sum of the initialization
delay, $t_{init}$ and the evaluation delay, $t_{eval}$. We note that this may
understate the advantage of STEM: as observed in
Section~\ref{sec:timing_example}, the $t_{init}$ phase can be overlapped with
the evaluation of the next gate since the delay penalty for initialization is
paid only once in the first logic stage.  Compared to STEM, the AND2, AND3, and
AND4 implementations using conventional ASL are, respectively, $1.6\times$,
$2\times$, and $3.4\times$ slower and $2.3\times$, $1.9\times$, and $6.9\times$
less energy-efficient.  The large delay and energy improvements for AND4 are
primarily due to a reduction from a total of seven input magnets with the
conventional ASL implementation to five input magnets for STEM.

From a delay and energy perspective, the MAJ3 gate is substantially similar to
the AND2 structure and sees the same level of improvement. A similar analysis
on MAJ5 gates shows that the conventional ASL implementation is $2.3\times$ slower
than STEM, while being $2.5\times$ less energy-efficient. As before, the 
advantage of STEM is larger for gates with more inputs.

\begin{figure}[ht]
\centering
	\subfigure[] {
		\includegraphics[width=7.5cm]{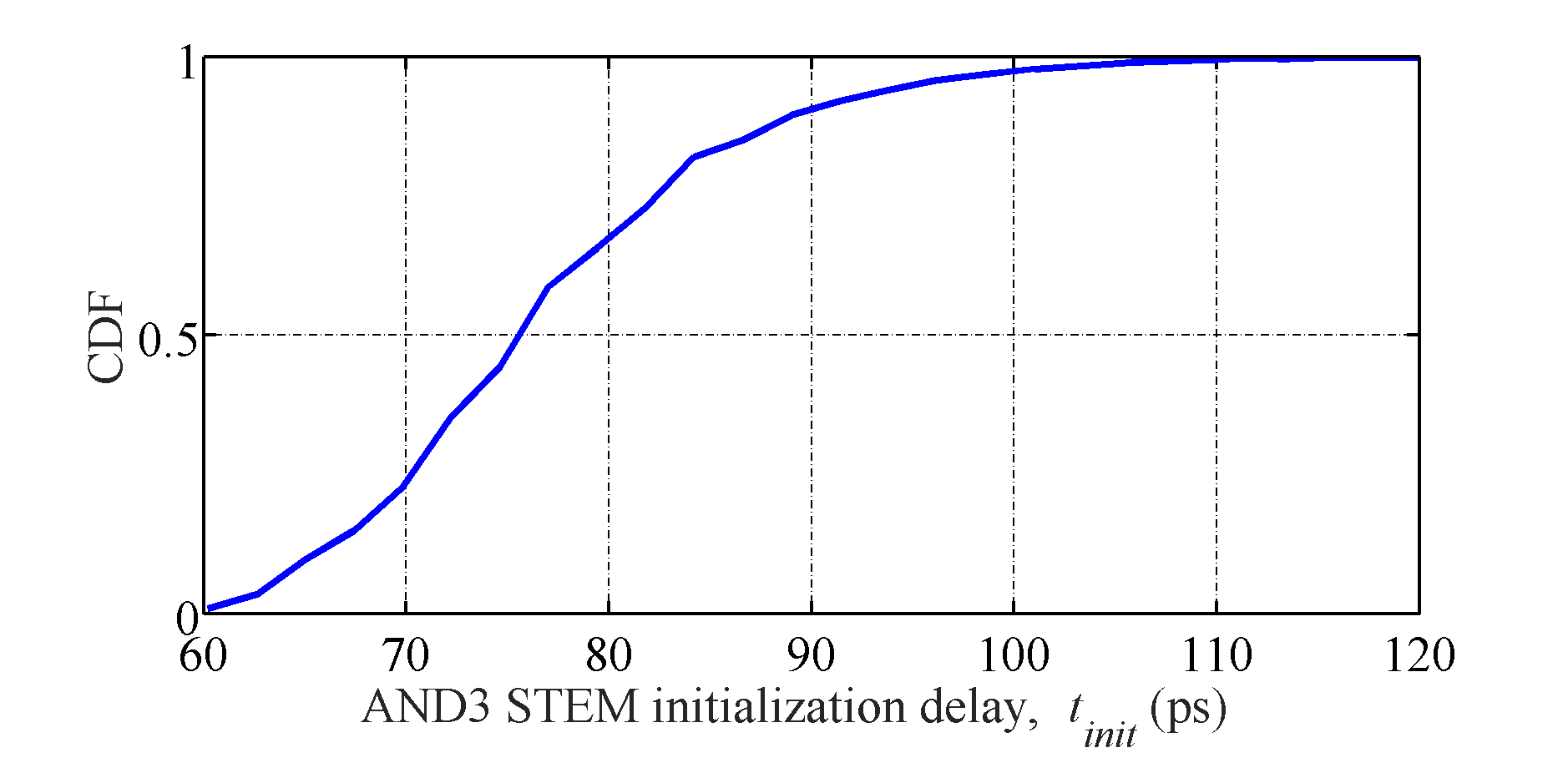}
	}
	\subfigure[] {
		\includegraphics[width=7.5cm]{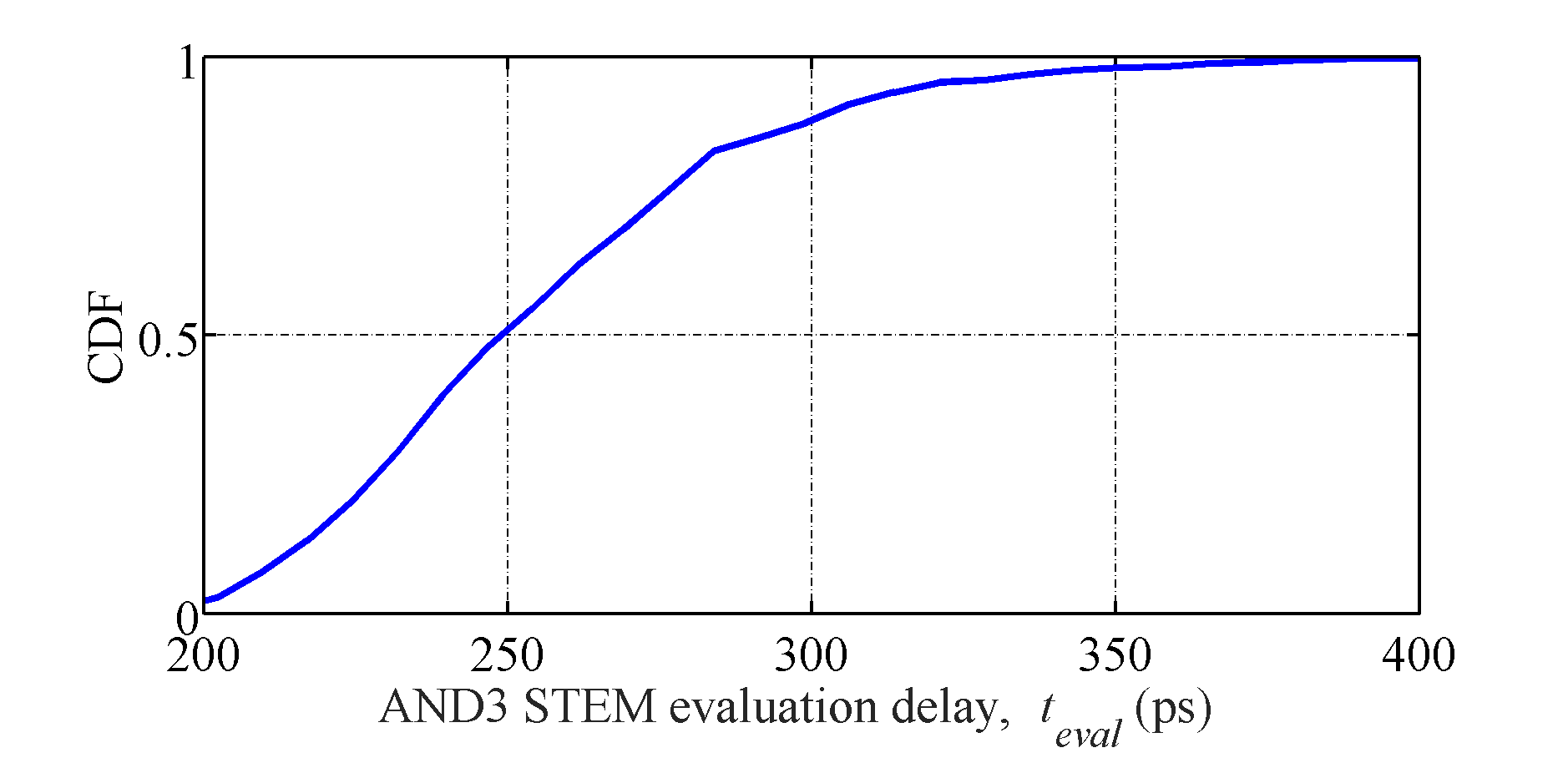}
	}
	\subfigure[] {
		\includegraphics[width=7.5cm]{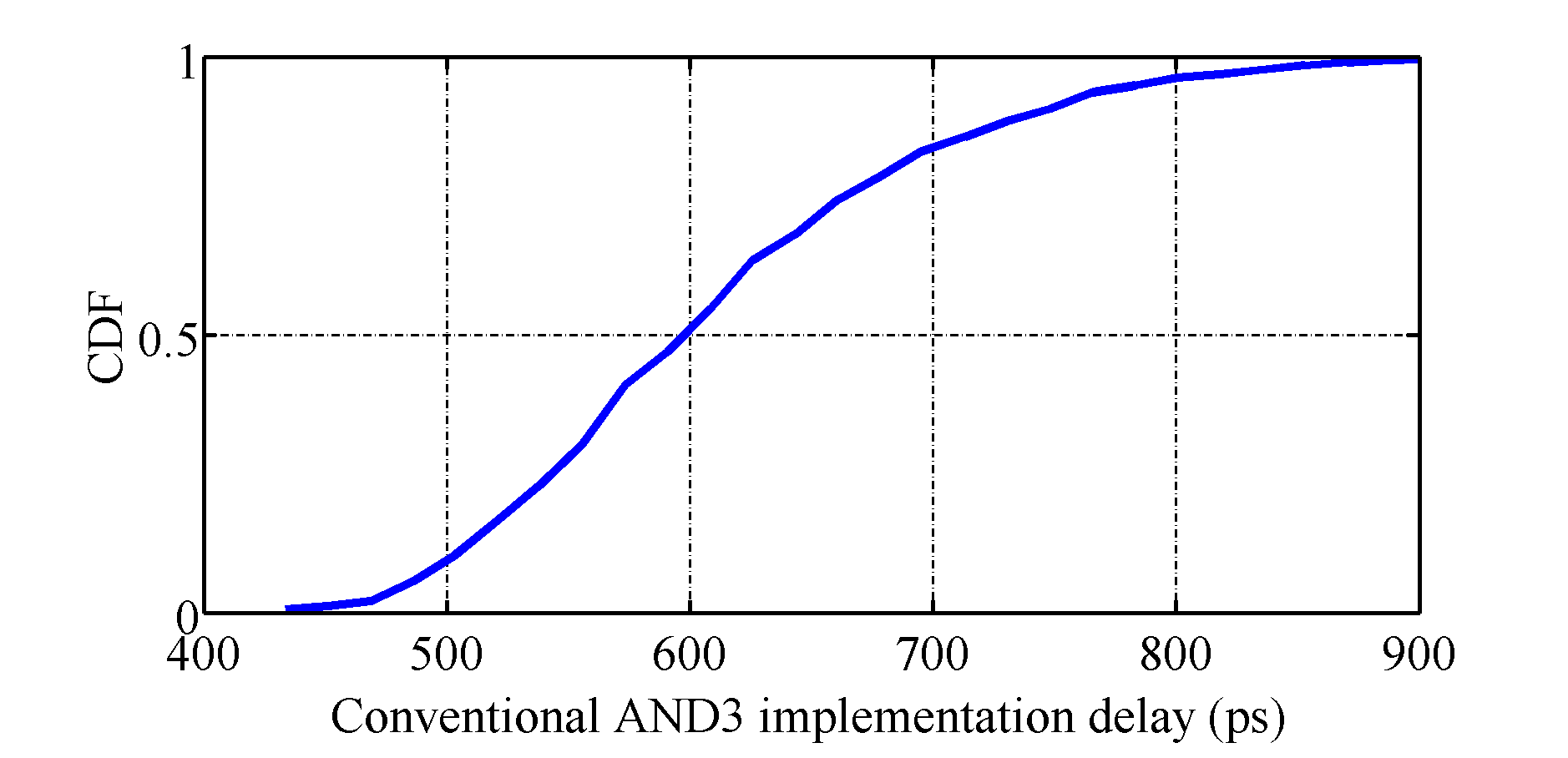}
	}
\caption{Cumulative distribution function of (a) the
initialization delay and (b) evaluation delay of the AND3 gate implemented 
using STEM technique, and (c) the delay of the AND3 gate using conventional implementation.}
\label{fig:AND3CDF}
\end{figure}

\subsection{Impact of thermal fluctuations on the switching delay}
\label{sec:thermal_fluc}

The switching delay of a ferromagnet in Equation~\eqref{eq:delay} is obtained as a
deterministic number by solving the Landau-Lifshitz-Gilbert-Slonczewski (LLGS) 
equation~\cite{Switching_energy_delay'11}. However, the switching process 
is indeterministic owing to the impact of random thermal fluctuations. 
Here, we study the variations in the switching delay with the example of an AND3 gate using 
the HSPICE model for stochastic LLGS~\cite{stochasticLLG} at room temperature. 
The AND3 circuit is modeled using the method described in
Section~\ref{sec:model}. The spin current, 
$I_s$, at the output magnet obtained by solving Equation~\eqref{eq:Gckt} is provided 
to the stochastic LLGS solver to obtain a distribution for the delay.

The cumulative distribution function (CDF) of the initialization
and the evaluation delay of the AND3 gate implemented using STEM scheme 
is shown in Figs.~\ref{fig:AND3CDF}(a) and~\ref{fig:AND3CDF}(b), 
respectively, while the corresponding CDF for the conventional
implementation is shown in Fig.~\ref{fig:AND3CDF}(c). 
The deterministic delay of Equation~\eqref{eq:delay} 
corresponds to the 99 percentile switching probability obtained from the stochastic LLGS 
solver. The 99 percentile point of the delay of the AND3 gate implemented using STEM
scheme is the sum of the 99 percentile point of the initialization and the 
evaluation delays. This ensures that the initialization of the output magnet of the AND3 is
complete before the evaluation phase begins. The AND3 STEM delay is thus
obtained as $\SI{438}{ps}$. The 99 percentile point
corresponds to a delay of $\SI{870}{ps}$ for the AND3 gate implemented
conventionally, which is approximately twice that of the STEM implementation. 
These numbers are consistent with the delays reported in
Fig.~\ref{fig:delay_energy}(a).

We make two important observations from Fig.~\ref{fig:AND3CDF}:
\begin{enumerate}
\item The delay distribution of the AND3 gate implemented with the
conventional scheme is broader compared to that of the AND3 gate
implemented with the STEM scheme. This result is consistent with the 
findings in~\cite{Butler}, which shows that the broadening of the delay
distribution occurs when the magnitude of the spin current that 
switches the output magnet is lowered. A lower spin current requires a
larger initial angle of the magnetization which in turn leads to a
larger time for the magnetization to achieve the 99 percentile point. 
In the case of the conventional AND3 implementation, the magnitude of the spin
current that switches the output magnet is less than that compared to
the STEM scheme as explained in detail in Sections~\ref{sec:majorityA}
and~\ref{sec:majorityB}, leading to a broader delay distribution. 
\item The stochastic nature of the magnet switching during the
initialization and the evaluation phase for STEM dictates the pulse
widths of the ``init'' and the ``eval'' signals. This is to ensure that
the initialization (evaluation) of the output magnet is complete when
the ``init'' (``eval'') signal is deasserted.  
\end{enumerate}

\subsection{Evaluating the five-magnet adder}
\label{sec:results_adder}

\begin{figure}[ht]
\centering
\includegraphics[height=5.75cm]{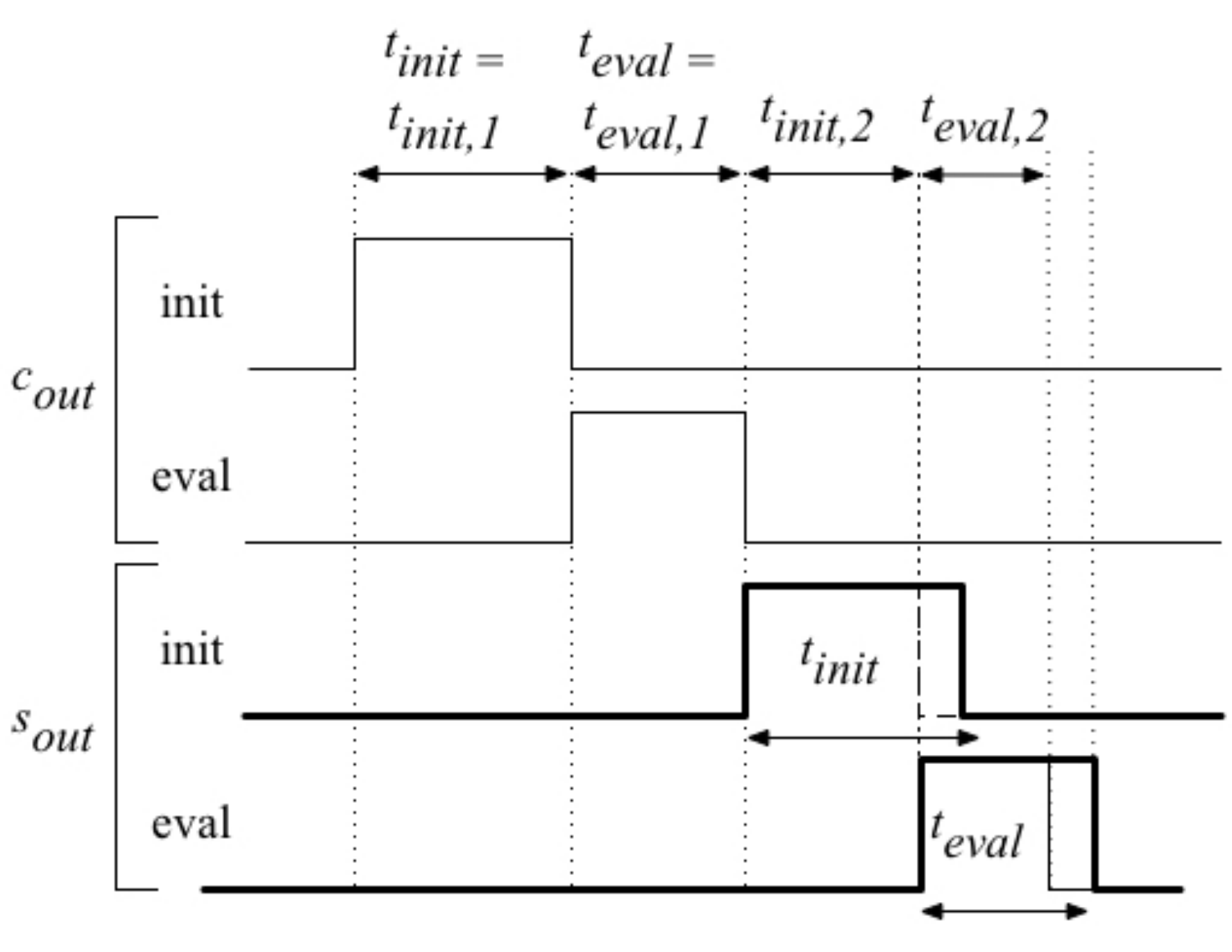}
\caption{A simplified timing scheme for the five-magnet adder.}
\label{fig:timing_adder}
\end{figure}

We show the impact of using STEM for implementing the five-magnet adder, as
described in Section~\ref{sec:adder}, and compare the delay and energy savings
with respect to the conventional ASL implementation. We model the adder
circuit in MATLAB using the method described in
Section~\ref{sec:model}. Using the notation
defined in Section~\ref{sec:adder}, the layout shown in
Fig.~\ref{fig:layout_adder}(b), and the simulation parameters in
Table~\ref{table:Parameters}, we find that $t_{init,1} =
\SI{425}{ps}$, $t_{eval,1} = \SI{370}{ps}$, $t_{init,2} = \SI{304}{ps}$, and
$t_{eval,2} = \SI{356}{ps}$. 

These results can be explained by referring
to the four-step operation of the full adder implemented with STEM presented in
Section~\ref{sec:adder}.
\begin{itemize}
\item The initialization of $c_{out}$ is performed using input $a$ in step
one. The requirement to achieve symmetry between the
input magnets, $a$, $b$, and $c_{in}$, for the evaluation of $s_{out}$ in step
four, prevents us from placing input magnet $a$ closer to $c_{out}$, for
a faster initialization in step one.
This results in a 1X current strength initializing $c_{out}$, while in step
two, a 2X current from input magnets, $b$
and $c_{in}$, from STEM MAJ3 implementation evaluates $c_{out}$. 
Therefore, $t_{init,1} > t_{eval,1}$. 
\item In step three, $c_{out}$
initializes $s_{out}$, while in step four, $s_{out}$ is evaluated as 
MAJ3($a$,$b$,$c_{in}$). Thus, there is no requirement for symmetry 
between the input magnets and $c_{out}$ for the
evaluation of $s_{out}$. This allows the placement of $s_{out}$ such that we
obtain a compact layout for the full adder with STEM scheme as shown in
Fig.~\ref{fig:layout_adder}(b). In comparison, the full adder layout
obtained using the conventional implementation occupies more area as
seen from Fig.~\ref{fig:layout_adder}(a). The evaluation of
$s_{out}$ in the conventional implementation as MAJ5($a$,$b$,$c_{in}$,
$c'_{out}$,$c'_{out}$) requires careful balancing of channel segment
lengths to achieve symmetry between the input magnets and $c_{out}$, 
thereby occupying more area.
\item Moreover, a 3X net current evaluates
$s_{out}$ in step four compared to the 2X current strength that 
evaluates $c_{out}$ in step two. Therefore, we obtain 
$t_{eval,1} > t_{eval,2}$. 
\item The channel length
between input magnet $a$ and $c_{out}$ is longer compared to the channel
length between $s_{out}$ and $c_{out}$. This results in
$t_{init,1} > t_{init,2}$.
\end{itemize}

We begin with the basic timing diagram of Fig.~\ref{fig:timing_adder1}.
The timing diagram with the relative pulse-widths of the initialization
and evaluation signals for the full adder implemented using the STEM
technique is shown in Fig.~\ref{fig:timing_adder}. 
Since it is
preferable to generate a single evaluation pulse that can be applied to the
gates, we set a safe value of $t_{init}~=~max(t_{init,1},t_{init,2})$
and $t_{eval}~=~\max(t_{eval,1},t_{eval,2})$ applied to both gates. We
therefore delay the ``eval'' signal to $s_{out}$ by $(t_{init,1}-
t_{init,2})$ time units, and extend the $t_{init,2}$ pulse duration to
equal $t_{init}$ time units. We also extend the ``eval'' signal to $s_{out}$
for an additional $(t_{eval,2} - t_{eval,1})$ units.  
The resulting timing diagram shown in Fig.~\ref{fig:timing_adder}
is identical to Fig.~\ref{fig:timing_adder1} with $t_{init} = t_{init,1}
= t_{init,2}$ and $t_{eval} = t_{eval,1} = t_{eval,2}$.
The total delay of the full adder implemented with the STEM scheme is
given by, $T_{adder,STEM} = 2(t_{init}+t_{eval})$.

\begin{table}[ht]
\centering
{\begin{tabular}{| l | r | r | r |}
\hline
\textbf{Adder implementation} & \textbf{Delay (ps)}  & \textbf{Energy
(pJ)} & \textbf{Area (nm$^2$)}\\
\hline
Conventional ASL & 2349 & 13.6 & 11250\\
\hline
STEM & 1590  & 6.3 & 9450\\
\hline
\end{tabular}
}
\caption{A comparison of the delay, energy, and area of the 
conventional ASL five-magnet adder implementation with STEM.}
\label{table:adder}
\end{table}

For the simulation parameters in Table~\ref{table:Parameters} and under the
spin circuit model in Section~\ref{sec:model}, the delay, energy, and
area for the full adder in the conventional ASL implementation and STEM 
are shown in Table~\ref{table:adder}. The full adder delay
using conventional ASL scheme, $T_{adder,conv}$, is the sum of the delays of
MAJ3 gate (to calculate $c_{out}$) and the MAJ5 gate (to calculate
$s_{out}$). Here, both MAJ3 and MAJ5 gates are implemented using the conventional
ASL scheme. Compared to the conventional implementation, we see that the
full adder implemented with STEM is $1.5\times$ faster and $2.2\times$ more 
energy-efficient, and provides a 16\% improvement in area.

\section{Conclusion}
\label{sec:conclusion}

In this work, we have proposed STEM, a novel two-phase method that leverages
the delay dependence of the device while implementing the majority logic. In
the first phase, the output is initialized to a preset value, while in the
second stage the inputs evaluate to switch the output under a time constraint.
We demonstrate this idea on standard cells built with ASL gates. We show that
an $n$-input AND gate which requires $(2n-1)$ inputs with conventional ASL can
now be implemented with just $(n+1)$ inputs. We show that STEM significantly
outperforms conventional ASL in terms of delay as well as energy: STEM ASL gates 
 are $1.6{\times}-3.4{\times}$ faster while being
$1.9{\times}-6.9{\times}$ more energy-efficient as compared to conventional ASL gates,
STEM ASL majority gates are $1.6{\times}-2.3{\times}$ faster and
$2.3{\times}-2.5{\times}$ more energy-efficient than conventional, and a
STEM ASL five-magnet full adder is $1.5\times$ faster, $2.2\times$ more
energy-efficient, and $1.2\times$ more area-efficient than its conventional counterpart. 
Further circuit-level and system-level optimizations are possible.  Like CMOS domino
logic~\cite{Harris97}, the STEM logic family is very amenable to pipelining and
we believe that many of the methods used to pipeline domino logic carry over to
STEM.

\bibliographystyle{IEEEtran}
\bibliography{ref}

\end{document}